\documentclass[10pt]{article}
\usepackage{amssymb,euscript,bbold}
\usepackage{amsfonts}
\usepackage{amssymb}
\usepackage{mathrsfs}
\usepackage{graphicx}
\usepackage{slashed}
\usepackage{amsthm}
\usepackage{amsmath}

\newtheorem{theorem}{Theorem}

\UseRawInputEncoding
\newcommand{\be}{\begin{equation}}
\newcommand{\ee}{\end{equation}}
\newcommand{\ba}{\begin{eqnarray}}
\newcommand{\ea}{\end{eqnarray}}

\newcommand{\op}[1]{\operatorname{#1}} 
\usepackage[table, svgnames]{xcolor}
\usepackage{bbold} 
\usepackage{hyperref}
\usepackage{mathtools}
\usepackage{listings}
\usepackage{longtable}
\usepackage{caption}
\usepackage{float}
\usepackage{pifont}
\newcommand{\xmark}{\ding{55}}%
\usepackage{tablefootnote}

\newcommand{\bbR}{{\mathbb R}}

\newcommand{\bbD}{{\mathbb D}}

\newcommand{\bbO}{{\mathbb O}}

\newcommand{\bbZ}{{\mathbb Z}}

\newcommand{\SO}{\operatorname{SO}}
\newcommand{\PSO}{\operatorname{PSO}}
\newcommand{\Sp}{\operatorname{Sp}}
\newcommand{\SU}{\operatorname{SU}}
\newcommand{\Or}{\operatorname{O}}
\newcommand{\Spin}{\operatorname{Spin}}

\renewcommand{\Re}{\mathrm{Re}}

\newcommand{\la}{\langle}
\newcommand{\ra}{\rangle}

\newcommand{\bpm}{\begin{pmatrix}}
\newcommand{\epm}{\end{pmatrix}}

\numberwithin{equation}{section}

\theoremstyle{remark}

\newtheorem{remark}{Remark}
\newtheorem{example}{Example}

\begin{document}
\input{epsf}

\begin{flushright}

\end{flushright}
\begin{flushright}
\end{flushright}
\begin{center}
\Large{On Classifying HyperK\"ahler Kummer 8-orbifolds}\\
\bigskip
\large{B.S. Acharya${}^{1}$}, 
\large{D.A. Baldwin${}^{1,2}$}\\
\smallskip\normalsize{\it
1. Abdus Salam International Centre for Theoretical Physics, Strada Costiera 11, 34151, Trieste, Italy}\\

and

{\it 2. Department of Physics, Kings College London, London, WC2R 2LS, UK}\\
\end{center}

\bigskip
\begin{center}
{\bf {\sc Abstract:}}

HyperK\"ahler spaces, including manifolds, orbifolds and conical singularities play an important role in superstring/$M$-theory and gauge theories as well as in differential and algebraic geometry. In this paper we provide hundreds of new examples of compact hyperK\"ahler orbifolds of Kummer type $T^8/G$, where $T^8$ is the maximal torus of the compact Lie group $E_8$ and $G$ a finite group of isometries whose holonomies form a subgroup of the Weyl group of $E_8$. We show that, out of all of these examples, the only orbifolds whose singularities have a known holomorphic symplectic resolution lead to manifolds diffeomorphic to the two currently known examples of compact hyperK\"ahler 8-manifolds. We also demonstrate that these methods can, when combined with theorems of Joyce, be extended to construct potentially new manifolds of $\op{SU}(4)$- and $\op{Spin}(7)$- holonomy. All of these examples give rise to new vacua of string/$M$-theory in two/three dimensions.
\end{center}

\newpage

\section{Introduction}

Einstein's equations with zero cosmological constant are equivalent to the existence of a metric $g(M)$ on the spacetime manifold $M$ whose Ricci tensor vanishes, $\op{Ric}(g(M))=0$. Such manifolds are called Ricci flat. They have played a ubiquitous role in both physics and mathematics for more than a century. In particular, in superstring/$M$-theory they are natural candidates as models for the extra dimensions of space and thus compact Ricci flat manifolds have been the subject of fairly intense study over the past four decades. At present the only known examples of compact, simply connected Ricci flat manifolds are those with special holonomy groups: $\op{Sp}(k), \op{SU}(m), G_2, \op{Spin}(7)$, corresponding to hyperK\"ahler, Calabi-Yau and exceptional holonomy respectively. Hence, special holonomy Ricci flat manifolds have played a particularly prominent role. Physically, the Ricci flat special holonomy groups are precisely those which preserve supersymmetry, which was the original motivation for studying Calabi-Yau manifolds ($\op{SU}(m)$-holonomy) in string theory \cite{candelas1985vacuum}. Presently, it is not known, apart from in real dimension four, if there are finitely or infinitely many diffeomorphism types of Ricci flat special holonomy manifolds in each dimension. This question is part of the motivation for this work, which can be viewed as a search for new hyperK\"ahler manifolds.

From a physical point of view, theories with larger supersymmetry groups are more tightly constrained than those without. From the special holonomy point of view, it is hyperK\"ahler manifolds which give rise to the largest supersymmetry groups and these will be the main subject of this paper. Part of the motivation is that,
at present, there are very few known examples of compact hyperK\"ahler manifolds. The current list, which we will state in the next section, comprises only two examples in each real dimension $4n$ \cite{beauville1983varietes} and an additional two exceptional examples in real dimension 12 and 20 \cite{OGrady1,OGrady2}. There have been no new examples found for three decades and it is not known if the current list is exhaustive or not.
Whilst Yau's proof of the Calabi conjecture applies to hyperK\"ahler manifolds, finding the explicit holomorphic symplectic K\"ahler manifolds to which it applies is the challenge.

The exceptional groups $G_2$ and $\op{Spin}(7)$ are also challenging Ricci flat special holonomy groups as there is no analogue of the Calabi conjecture and the best available theorems are those of Joyce \cite{joyce1996compact,joyce1996compactII,joyce1996compactspin7, JoyceBook} which are based on demonstrating that if one is sufficiently close to a special holonomy metric, then under suitable circumstances one can perturb to a genuine Ricci flat metric. Joyce applied these results to go on to construct the first compact examples of $G_2$ and $\op{Spin}(7)$ holonomy manifolds by a generalised Kummer construction, where one attempts to resolve the singularities in a finite quotient, $T^7/G$ or $T^8/G$, by removing small neighbourhoods of the singular set and gluing in suitably small model manifolds and Ricci flat metrics and perturbing.
One of the key aspects of Joyce's work is that one has to carefully choose the finite group actions on $T^n$ in order that the singular sets are suitably mild enough, typically consisting of codimension four $ADE$ singularities which ideally do not intersect each other. This is achieved by carefully choosing the rotations {\it and} translations in $G$ since, without translations, all components of the singular set of $T^n/G$ intersect at the origin whereas with them, certain elements of the group act freely on the fixed locus of other elements. 

The idea behind this paper is to try to construct compact hyperK\"ahler manifolds using a similar strategy. To this end we consider the Weyl group $W(E_8)$ of $E_8$ acting on $T^8 \cong T(E_8)$, the maximal torus of $E_8$. First we classify all of the subgroups of $W(E_8)$ preserving a hyperK\"ahler structure on $T^8$. This leads to 139 model hyperK\"ahler orbifolds $T^8/H$, with the elements of $H$ acting as rotations in $\op{O}(8)\times\{\mathbb{1}\}\subset \op{O}(8)\ltimes \mathcal{T}(T^{8})\cong \mathcal{E}_{8}$ where $\mathcal{E}_{8}$ is the Euclidean group i.e. the translations are trivial.
We then go on to show that, no matter which translations are included compatible with the $H$-action, it is impossible to generate singular sets which have crepant resolutions, apart from those diffeomorphic to the previously known examples. \emph{This proves that no new examples of compact hyperK\"ahler 8-manifolds can be constructed with this method.}

In order for the hyperK\"ahler orbifolds we consider to admit a  desingularisation to a compact hyperk\"ahler manifold, the singularities occuring in the singular set must admit a \emph{symplectic resolution}. The classification of such singularities is not complete but it is thought that few, if any, more examples will be added to the current known list, see \cite{bellamy2022towards, bellamy2023parabolic} for details.  Generically, hyperK\"ahler orbifolds will have some singularities that are known not to admit symplectic resolutions. 
We will call such singularites \emph{non-resolvable symplectic (NRS) singularities}. They may also contain singularities for which a symplectic resolution is not known to exist, and these will be called \emph{undecided symplectic (US) singularities}. These are described in detail in \cite{bellamy2022towards, bellamy2023parabolic} and examples will appear throughout this paper.

The maximal torus of $E_{8}$ was chosen because it has a huge symmetry group given by $W(E_{8})$ and so it can provide many potentially interesting orbifolds through quotients by finite subgroups. 
The subgroups of $W(E_{8})$ that preserve a hyperK\"ahler structure can be classified and will be described in section \ref{sec: Robert's classification}. 
It turns out there are three classes of such subgroups, which we call $G_{i}$  ($i=1,2,3$). For every subgroup $H$ of $G_1, G_2$ or $G_3$ one can consider including all possible translations compatible with the $H$ action on $T^8$ giving a group $H'$ isomorphic to $H$ but with a different action\footnote{In the language of crystallographic groups, there can be more than one space group for a given point group $H$.} on $T^8$. For any such hyperK\"ahler orbifold, $T^8/H'$, there are three possibilities: $(i):$ $T^8/H'$ admits a symplectic resolution to a compact hyperK\"ahler manifold; $(ii):$ $T^8/H'$ has at least one NRS singularity and hence can not be resolved to a compact hyperK\"ahler manifold; $(iii):$ $T^8/H'$ has at least one US-singularity and no NRS-singularities and we currently are unable to show that it can or cannot admit a symplectic resolution.

Our main results are given in Tables \ref{tab:table 1}, \ref{tab:table 2} and \ref{tab:table 3}. Each line in the table is a subgroup $H$. This gives 139 compact hyperK\"ahler 8-orbifolds, of which one is Type $(i)$, 138 are Type $(ii)$ and none are Type $(iii)$. We further show that every choice of $H'$ for the Type $(i)$ case leads to a Type $(ii)$ orbifold.
For the 138 Type $(ii)$ examples we are able to show that for every choice of $H'$ we can never remove all of the NRS-singularities and, hence the orbifolds will remain Type $(ii)$.

Although we were unable to produce new hyperK\"ahler manifolds, this construction does provide many new compact hyperK\"ahler orbifolds which may still be of interest. Furthermore, the methods in this paper can certainly be applied to study higher dimensional hyperK\"ahler orbifolds as well as other holonomy groups. In fact we show that one can construct new Calabi-Yau manifolds using this approach and provide a pair of examples at the end of the paper.

In section \ref{sec: hyperkahler geom}, we review some basic properties of compact hyperK\"ahler manifolds as well as details about symplectic singularities and their resolutions. In section \ref{sec: Robert's classification}, we describe the classification of all the subgroups of $W(E_{8})$ preserving a hyperK\"ahler structure and then explicitly represent these subgroups as matrix groups acting on $\mathbb{R}^{8}$. We then provide the tables detailing all 139 of the model hyperK\"ahler orbifolds. In section \ref{ref: studying singular sets}, we consider the singular sets of these orbifolds for all choices of translations and describe the method by which we can determine if an orbifold is non-resolvable. 

\section{HyperK\"ahler Geometry}\label{sec: hyperkahler geom}

\subsection{Compact hyperK\"ahler manifolds}

The following can found in, for example, \cite{huybrechts1997compact}. Compact hyperK\"ahler manifolds can be viewed from the Riemannian point of view or from the complex geometric point of view, where they are often known as irreducible holomorphic symplectic manifolds. From the former viewpoint, they are $4n$-dimensional compact Riemannian manifolds $(M,g)$ with holonomy group $\op{Hol}(g)=\op{Sp}(n)$. From the latter viewpoint, they are compact K\"ahler manifolds $X$ that are simply connected and admit a $(2,0)$-form that is closed and non-degenerate on all of $X$, they then necessarily have even complex dimension $2n$. To pass from the former to the latter viewpoint is straightforward: a hyperK\"ahler manifold is equipped with three integrable complex structures $I, J, K$ that are K\"ahler with respect to the hyperK\"ahler metric $g$, call the K\"ahler forms $\omega_{I,J,K}$. One can then take, for instance, the K\"ahler manifold $(M,I)$ which admits a holomorphic 2-form that is closed and non-degenerate on all of $M$ given by $\Omega = \omega_{J} + i \omega_{K}$. Furthermore, $(M,I)$ is simply-connected by virtue of being a compact Riemannian manifold with holonomy $\op{Sp}(n)$ and also $(M,I)$ cannot be written as the product of lower-dimensional complex manifolds else $g$ would not have irreducible holonomy. To go from the latter to the former viewpoint is slightly more involved and the details can be found in $\cite{huybrechts1997compact, beauville1983varietes}$.

As mentioned in the introduction, in contrast to, say, Calabi-Yau manifolds, examples of compact hyperK\"ahler manifolds are very few and far between and no new examples have been found for nearly 30 years. In real dimension 4 the only example is a $K3$ surface. The full known list for real dimension greater than 4 is the following,

\begin{itemize}
    \item[(i)] The \emph{Hilbert scheme of $n$ points} $\op{Hilb}^{n}(X)$ where $X$ is a $K3$ surface \cite{beauville1983varietes}. This manifold has real dimension $4n$ and $b_{2}(\op{Hilb}^{n}(X)) = 23$ for $n>1$.
    \item[(ii)] The \emph{generalised Kummer variety} $\op{K}_{n}(X)$, where $X$ is a real 4-dimensional torus \cite{beauville1983varietes}. This has real dimension $4n$ and $b_{2}(\op{K}_{n}(X)) = 7$ for $n>1$.
    \item[(iii)] A real dimension 20 example due to O'Grady \cite{OGrady1}. It has $b_{2} = 24$ \cite{de2021hodge}.
    \item[(iv)] A real dimesnion 12 example due to O'Grady \cite{OGrady2}. It has $b_{2} = 8$.
\end{itemize}
We will elaborate slightly on each of these examples. The Hilbert scheme $\op{Hilb}^{n}(X)$ can be understood as a crepant resolution of the symmetric product $X^{n}/S_{n}$. For $n=2$ this resolution can be seen explicitly as a blow-up of the $A_{1}$-singularities lying along the diagonal surface in $X^{2}/\mathbb{Z}_{2}$. When $X$ is a $K3$ surface or a torus $T^{4}$ it is true that $\op{Hilb}^{n}(X)$ is holomorphic symplectic however only when $X$ is a $K3$ surface is $\op{Hilb}^{n}(X)$ also irreducible \cite{beauville1983varietes}. When $X$ is $T^{4}$, $\op{Hilb}^{n}(X)$ has a finite cover isomorphic to $K_{n-1}(X)\times X$. We can see $K_{n-1}(X)$ as the kernel of the map,
\begin{equation}
    \op{Hilb}^{n}(X)\overset{\pi}{\rightarrow}X^{n}/S_{n} \overset{\Sigma}{\rightarrow} X
\end{equation}
where $\Sigma$ is the map which takes the sum of the $n$ points. Note that $K_{1}(T^{4})$ is a $K3$ surface obtained from the Kummer construction, hence the name `generalised kummer variety'. The examples due to O'Grady were obtained as crepant resolutions of the singular moduli space of certain sheaves on $K3$ surfaces.

\subsection{Symplectic resolutions}\label{sec: symp resolutions}
Let $G\subset \op{Sp}(n)$ be a finite subgroup and $V$ be a symplectic representation of $G$. Then $V/G$ is a \emph{symplectic singularity} and admits a 2-form that is closed and non-degenerate on all of the smooth part of $V/G$. A desingularisation $\widetilde{V/G}\rightarrow V/G$ is called a \emph{symplectic resolution} if the pull back of the 2-form on $V/G$ is closed and non-degenerate on all of $\widetilde{V/G}$. The current list of  singularities $V/G$ known to admit a symplectic resolution is the following \cite{bellamy2013new}:
\begin{itemize}
    \item[(i)] $G = S_{n+1}$ acting on $V = W\oplus W \cong \mathbb{C}^{2n}$ where $W = \left\{x \in \mathbb{C}^{n+1}\mid \sum x_{i}=0\right\}$ and $S_{n+1}$ acts by permuting $x_{1}$,..., $x_{n}$.
    \item[(ii)] $G = H\wr S_{n}$ (where $\wr$ denotes the wreath product) acting on $V = \mathbb{C}^{2n}$ in the following way,
    \begin{equation}
        \frac{\mathbb{C}^{2n}}{G} = \op{Sym}^{n}\left(\mathbb{C}^{2}/H\right)
    \end{equation}
    where $H\subset \op{SU}(2)$ is a finite subgroup acting in the fundamental representation and $S_{n}$ permutes the factors.
    \item[(iii)] $G= \mathbb{T}$, the binary tetrahedral group, acting in the following way: take the standard irreducible 2-dimensional representation on $\mathbb{C}^{2}$ given by $\mathbb{T}\subset \op{SU}(2)$, i.e. with generators given by,
    \begin{equation}
        g_{1} = \frac{1}{2}\begin{pmatrix}
            -1+i&1+i\\-1+i&-1-i
        \end{pmatrix},\quad g_{2} = \frac{1}{2}\begin{pmatrix}
            -1+i&-1-i\\1-i&-1-i
        \end{pmatrix}
    \end{equation}
    and take the tensor product with one of the two non-trivial 1-dimensional representations (e.g. where $g_{1}$ is given by multiplication by $\op{exp}(2\pi i/3)$), then take the direct sum of this representation with its dual representation. In other words, take the generators to be,
    \begin{equation}
        \hat{g}_{1} = \begin{pmatrix}
            e^{\frac{2\pi i}{3}}\,g_{1} & \\ & e^{-\frac{2\pi i}{3}}\,g_{1}^{*} 
        \end{pmatrix},\quad \hat{g}_{2} = \begin{pmatrix}
            e^{\frac{2\pi i}{3}}\,g_{2} & \\ & e^{-\frac{2\pi i}{3}}\,g_{2}^{*} 
        \end{pmatrix}
    \end{equation}
    \item[(iv)] $G = Q_{8}\circ D_{4}$, the central product of the quaternion group with the dihedral group of order 8, acting in the tensor product representation of the irreducible 2-dimensional representation of $Q_{8}$ given by,
    \begin{equation}
        I = \begin{pmatrix}
            0&-1\\1&0
        \end{pmatrix}, \quad J = \begin{pmatrix}
            i&0\\0&-i
        \end{pmatrix}, \quad K = \begin{pmatrix}
            0&i\\i&0
        \end{pmatrix}
    \end{equation}
    with that of $D_{4}$ given by,
    \begin{equation}
        r = \begin{pmatrix}
            0&-1\\1&0
        \end{pmatrix},\quad f = \begin{pmatrix}
            1&0\\0&-1
        \end{pmatrix}
    \end{equation}
    For example, one can take the matrices,
    \begin{equation}
        I\otimes r,\quad I\otimes f,\quad J\otimes r,\quad K\otimes r
    \end{equation}
    as generators.
\end{itemize}
This is not yet a complete classification. In \cite{bellamy2013new} the authors state that they suspect there are few (if any) other examples remaining to be discovered, see there for more discussion on this point. In the more recent works \cite{bellamy2022towards, bellamy2023parabolic}, the authors reduce the open cases from an infinite series of groups in complex dimension 4 (so-called symplectic primitive, complex imprimitive reflection groups, see \S 3 of \cite{cohen1980finite}) and 13 exceptional examples\footnote{Matrix representations (for GAP or MAGMA implementation) of all 13 of the symplectic primitive, complex primitive reflection groups are provided on the Github page of one of the authors of the above-cited works, \url{https://github.com/joschmitt/Parabolics}.} (symplectic primitive, complex primitive reflection groups, see \S 4 table III of \cite{cohen1980finite}) to just 39 examples from the infinite family and 6 exceptional cases. Several of the 45 open cases appear in the examples we consider, they are marked by stars in the tables in \ref{sec: list of subgroups}.

It is not expected by the authors of the above-cited works that any of the remaining open cases will admit a symplectic resolution.

\section{Quotients of the Maximal Torus of $E_{8}$}\label{sec: Robert's classification}
\subsection{Classification of subgroups of $W(E_{8})$ preserving a hyperK\"ahler structure}

In this section we describe a classification of the subgroups of the Weyl group of $E_{8}$ that preserve a hyperK\"ahler structure on $T^{8}$, the maximal torus of $E_{8}$.

We make use of the fact that $\Spin(8)$ has three distinct $8$-dimenionsal representations $\pi_i:\Spin(8)\to\SO(V_i)$ for $i=1,2,3$ so that the subgroups that preserve an $\Sp(2)$-structure of one orientation in $V_1$ correspond to the subgroups that are the identity on some $3$-plane in $V_2$ while the subgroups that preserve an $\Sp(2)$-structure of the other orientation in $V_1$ correspond to the subgroups that are the identity on some $3$-plane in $V_3$.

The root system of $E_8$ in $\bbR^8 = \bbO$ is usually described as
$$
\Phi_0 = \{\,{}\pm e_i \pm e_j\,|\,1\le i< j\le 8\,\} \cup \{\,\tfrac12(\epsilon_1\,e_1 +\cdots + \epsilon_8\,e_8)\,|
\,\epsilon_i = \pm 1,\ \epsilon_1\cdots\epsilon_8 = 1\,\}.
$$
Note that $|\Phi_0| = 240$ and that every $u\in\Phi_0$ satisfies $|u|^2 = 2$.  

For our purposes, though, it will be much more convenient 
to invoke the ring of integral octonions $\bbD\subset\bbO$ 
of Dickson and Coxeter (see appendix \ref{sec: int octs}), for then we can simply let 
$$
\Phi = \bbD^\times = \{\ x\in\bbD\ |\ |x|^2 = x\bar x = 1\ \}.
$$
$\Phi$ is also a root system of type $E_8$.
Since $|xy| = |x|\,|y|$ for $x,y\in\bbO$, it follows that $\Phi$ 
is a multiplicative subset of $\bbD$, one that includes the multiplicative unit~$1$.  \emph{However, it must be emphasized that, because $\bbD^\times$ 
is not associative, $\Phi$ does not form a group under multiplication.}

The Weyl group $W(E_{8})$  of $E_8$ is generated by the reflections~$\rho_u$ 
for $u\in \Phi$, which have the formula
$$
\rho_u = -L_u\circ R_u \circ c = L_u\circ R_u \circ \rho_1 
= \rho_1\circ L_{\bar u}\circ R_{\bar u}
$$
where $c = -\rho_1$ is octonionic conjugation $c:\bbO\to\bbO$.  (Note that
since $\rho_1\in W(E_{8})$ and $-I_8\in W(D_8)\subset W(E_{8})$, we have $c\in W$.)
In particular, since ${\rho_1}^2 = 1$, it follows that 
$$
\rho_u\circ\rho_v = L_u\circ R_u \circ L_{\bar v}\circ R_{\bar v}\,,
$$
a formula that will be important below.

The order of $W(E_{8})$  is known to be $696,729,600= 2^{14}\cdot3^5\cdot5^2\cdot7$.  Hence, the order of $W^+ = W(E_{8})\cap \SO(8)$ is $348,364,800 =2^{13}\cdot3^5\cdot5^2\cdot7$.  $W^+$ is the group of linear transformations that are a product of an \emph{even} number of the 120 distinct reflections $\{\,\rho_u\ |\ u\in \Phi\,\}$.  The center of $W^+$ is $\{\pm I_8\}$, and the
quotient $W^+/\{\pm I_8\}\subset\SO(8)/\{\pm I_8\} = \PSO(8)$ is a simple group of order $174,182,400 = 2^{12}\cdot3^5\cdot5^2\cdot7$.

Our goal is to find a way to locate subgroups of $W^+$ that preserve a hyperK\"ahler structure on $\mathbb{O}$.  We will do this by exploiting
triality in $\SO(8)$ (see appendix \ref{sec: trialilty} to fix notation 
and the basic properties of triality).

Let $\tilde W^+ = {\pi_1}^{-1}(W^+)\subset\Spin(8)$ 
denote the `binary even Weyl group' of $E_8$.  Since $W^+\subset\SO(8)$
contains the center $\{\pm I_8\}$ of $\SO(8)$, it follows that $\tilde W^+$
contains the center of $\Spin(8)$.  In particular, $\tilde W^+$ 
contains $(I_8,-I_8,-I_8)$.  Because of the identity 
$\rho_u{\circ}\rho_v = L_u{\circ}R_u{\circ}L_{\bar v}{\circ}R_{\bar v}$, 
it follows that ${\pi_1}^{-1}(\rho_u\circ\rho_v)\in\tilde W^+$ 
consists of the two elements 
$(\rho_u{\circ}\rho_v, \pm L_{\bar u}L_{v},\pm R_{\bar u}R_{v})$ 
for any $u,v\in\Phi$.  
Hence, the set
$$
\Gamma 
= \{\,(\rho_u{\circ}\rho_v,L_{\bar u}L_{v},R_{\bar u}R_{v})\,|\,u,v\in\Phi\,\}
\subset\Spin(8).
$$ 
generates $\tilde W^+$. 

It follows that $\pi_2(\tilde W^+)\subset\SO(8)$ 
is generated by $\{\,L_{u} L_v\,|\,u,v\in\Phi=\bbD^\times\,\}$ 
and preserves $\Phi = \bbD^\times$.  Hence it lies in $W^+$ 
(which is the subgroup of $\SO(8)$ that preserves $\Phi$). 
Since $\pi_2(\tilde W^+)$ and $W^+$ have the same cardinality, 
it follows that $\pi_2(\tilde W^+) = W^+$. 
Similarly, $\pi_3(\tilde W^+) = W^+$.  

However, if $\Gamma\subset W^+$ is a proper subgroup of $W^+$,
then it need not be the case that 
$\Gamma_2=\pi_2\bigl(\pi_1^{-1}(\Gamma)\bigr)$ 
and $\Gamma_3=\pi_3\bigl(\pi_1^{-1}(\Gamma)\bigr)$ 
are isomorphic to $\Gamma$.

\begin{example}
For example, suppose that $\Gamma$ is the subgroup of $W^+$ 
that fixes an element $u\in\Phi = \bbD^\times$.  
Since $W^+$ acts transitively on $\Phi$, we may as well take $u=1$.
Now, $\Gamma\subset\SO(1^\perp)$ consists of symmetries
of the root system $1^\perp\cap \Phi$, which is known to be 
isomorphic to the root system of $E_7$ and is generated by
the $63$ reflections $\rho_v$ for $v\in 1^\perp\cap \Phi$,
i.e., the Weyl group of $E_7$, which is known to be the
full isometry group of the root system of $E_7$.  
Consequently, $\Gamma$ is the group generated by~$\rho_v{\circ}\rho_w$
for $v,w\in 1^\perp\cap \Phi$.  

It follows that $\Gamma_2$ is the group generated by $L_vL_{w}$ 
for $v,w\in 1^\perp\cap\Phi$. Via the identification $\Phi=\bbD^\times$, 
it is easy to show that $\Gamma_2$ acts transitively on~$\Phi$.
Moreover, since $-I_8\not\in\Gamma$, 
it follows that $\pi_2:\pi_1^{-1}(\Gamma)\to\Gamma_2$ is injective
and hence an isomorphism.  Moreover, $\Gamma_2$ acts preserving
a $\Spin(1^\perp)$-structure, i.e., a $\Spin(7)$-structure, on $\bbO$
(in addition to preserving the lattice $\bbD\subset\bbO$). 
\end{example} 

\subsubsection{Subgroups preserving special structures}
Generalizing the above example, we are going to be interested in
subgroups $\Gamma$ of the even Weyl group that act as the identity on 
a subspace of $\bbO$ of dimension $1$ (respectively, $2$, or $3$), 
because these are the conditions that ensure that the corresponding group
$\Gamma_2$ or $\Gamma_3$ acts preserving the lattice $\bbD$ 
and a $\Spin(7)$-structure (respectively, a $\SU(4)$-structure
or a $\Sp(2)$-structure, since $\Spin(6) = \SU(4)$ and $\Spin(5) = \Sp(2)$).

Suppose that a subgroup $\Gamma\subset W^+$ 
acts as the identity on a subspace $S\subset\bbO$.
Fix a base $B\subset\bbD^\times$ for the root system
and let $F = \{f_1,\ldots,f_8\}\subset \bbD^\times$ 
be the corresponding set of fundamental weights, 
which span the extreme edges of the corresponding 
fundamental Weyl chamber.  Let $s\in S$ be
a `generic' element.  After conjugation in $W^+$, we can
assume that $s$ lies in the fundamental Weyl chamber.  
Writing $s = s_1\,f_1 + \cdots + s_8\,f_8$ with $s_i\ge0$,
we know that any element of $W^+$ that fixes $s$ must
fix every $f_i$ where $s_i>0$.  Thus, $\Gamma$ must fix
each such $f_i$.  The result is that, if $\Gamma$ fixes
the elements of a subspace $S$, then it fixes the elements
of a (possibly larger) subspace $\bar S$ that is spanned by
a set of fundamental weights.  

Consequently, in classifying the subgroups of $W^+$ 
that leave the elements of some subspace $S\subset\bbO$ fixed, 
it suffices to consider only the subspaces $S$ 
that have a basis drawn from $\Phi = \bbD^\times$.  
\emph{In this case, $S\cap\Phi$ is a root system 
of rank equal to $\dim S$.}  

Now, the possible root systems $\Psi\subset\Phi$ of ranks $1$, $2$, 
or $3$, are easily listed.  Since all the roots in $\Phi$ have
the same length, it follows that the only possible \emph{irreducible}
root systems to be considered are $\mathrm{A}_1$, $\mathrm{A}_2$, 
and $\mathrm{A}_3$.
This means that we need only consider subsystems $\Phi$ of type
\begin{itemize}
\item Rank 1:  $\mathrm{A}_1$
\item Rank 2:  $\mathrm{A}_2$ or $\mathrm{A}_1\times \mathrm{A}_1$
\item Rank 3:  $\mathrm{A}_3$, $\mathrm{A}_2\times \mathrm{A}_1$, 
or $\mathrm{A}_1\times \mathrm{A}_1\times \mathrm{A}_1$.
\end{itemize}

Since $W^+$ acts transitively on $\Phi$, it follows that all the
root systems of type $A_1$ are conjugate to $\{1,-1\}$.  
As was already explained in the above Example, the subgroup of $W(E_{8})$
that preserves $1$ is the Weyl group of the root system $1^\perp\cap\Phi$,
i.e., the Weyl group of $\mathrm{E}_7$.  

For $\mathrm{A}_1\times \mathrm{A}_1$, we can take $S$ to be spanned by $1$ 
and $a\in 1^\perp\cap\Phi$.  Since the Weyl group of $\mathrm{E}_7$ acts
transitively on the root system, it does not matter 
which $a\in 1^\perp\cap\Phi$ we pick.  In this case, it is known
(and easy to see) that $1^\perp\cap a^\perp\cap\Phi$ 
is a root system of type $\mathrm{D}_6$.  
Moreover, the subgroup of $W(E_{8})$  that fixes $1$ and $a$ is easily
seen to be the Weyl group of $1^\perp\cap a^\perp\cap\Phi$.
Hence, the subgroup $\Gamma\subset W^+$ 
is the even Weyl group of $1^\perp\cap a^\perp\cap\Phi$ 
(which has type $\mathrm{D}_6$), a group of order $11,520= 2^8\cdot3^2\cdot5$.

Since the Weyl group of $\mathrm{D}_6$ acts transitively on the roots,
it follows that, for $\mathrm{A}_1\times \mathrm{A}_1\times \mathrm{A}_1$, 
we can take $S$ to be spanned by $1\in\Phi$, $a\in 1^\perp\cap\Phi$,
and $b\in 1^\perp\cap a^\perp\cap\Phi$.  It then turns out that
$1^\perp\cap a^\perp\cap b^\perp\cap\Phi$ is a root system of
type $\mathrm{A}_1\times \mathrm{D}_4$.  The subgroup $\Gamma\subset W^+$ 
is therefore isomorphic to the Weyl group of $\mathrm{D}_4$, 
a group of order $2^3\cdot 4! = 192$.  

For $\mathrm{A}_2$, one can show that $W(E_{8})$  acts transitively on the 
$2$-planes $S\subset\bbO$ such that $S\cap\Phi$ is a root
system of type $\mathrm{A}_2$.  Moreover, in this case, $S^\perp\cap\Phi$
is a root system of type $\mathrm{E}_6$.  Hence the subgroup $\Gamma$
that fixes $S$ is isomorphic to the Weyl group of $\mathrm{E}_6$.

For $\mathrm{A}_2\times \mathrm{A}_1$, one can show that, 
since the Weyl group of $\mathrm{E}_6$ acts transitively 
on the roots of $\mathrm{E}_6$, we can 
take the subspace spanned by an $\mathrm{A}_2$ root system 
plus an $\mathrm{A}_1$
that lies in the orthogonal $\mathrm{E}_6$ root system.  This gives 
a space $S$ of dimension $3$ such that $S^\perp\cap\Phi$ is
a root system of type $\mathrm{A}_5$.  (This corresponds to the 
Wolf space $\mathrm{E}_6/(\SU(6)\cdot\SU(2))$.)  Thus, 
the $\Gamma\subset W^+$ in this case will be isomorphic 
to the group $A_6\subset S_6$ that consists of even permutations 
of $6$ elements.

Finally, for $\mathrm{A}_3$, we can take advantage of the
fact that $\mathrm{E}_8$ contains $\SO'(16)$, which contains a
double cover of $\SO(6)\times\SO(10)$, 
implying that the root system of $\mathrm{E}_8$ contains
a root system $\mathrm{D}_3\times \mathrm{D}_5 
= \mathrm{A}_3\times \mathrm{D}_5$.  It is not hard to show
that $W(E_{8})$ acts transitively on the set of subsystems of type
$\mathrm{A}_3\times \mathrm{D}_5$, so all of these cases are
conjugate.  Consequently, in this case, we will have a $3$-dimensional
subspace $S\subset\bbO$ such that $S^\perp\cap\Phi$ is a root
system of type $\mathrm{D}_5$.  In this case, the subgroup $\Gamma$
of $W^+$ will be the even Weyl group of $\mathrm{D}_5$, a group of 
order $(2^4\cdot 5!)/2 = 960$.  

In particular, each of these three cases in which $S$ has dimension $3$
gives rise to a discrete subgroup $\Gamma_2\subset W^+$ that preserves
an $\Sp(2)$-structure on $\bbO$ (as well as the lattice $\bbD$).  
These groups have order $384$ 
(when $S\cap\Phi$ is a root system 
of type $\mathrm{A}_1\times \mathrm{A}_1\times \mathrm{A}_1$), $720$ (when $S\cap\Phi$ is a root system 
of type $\mathrm{A}_2\times \mathrm{A}_1$) or $1920$ (when $S\cap\Phi$ is a root system of type
$\mathrm{A}_3$) and we denote them $G_{1}$, $G_{2}$ and $G_{3}$ respectively.

\subsection{Explicit matrix representations}\label{sec: mat reps}
Following from the previous general considerations, we can now find explicit matrix representations of the three groups $G_{1}$, $G_{2}$ and $G_{3}$. Let $M_{\text{gen}}$ denote the `generator matrix' of the lattice, i.e. the matrix whose columns are the basis vectors of the lattice written in the standard basis of $\mathbb{R}^{8}$. The basis vectors in this case are given by a choice of simple roots for $E_{8}$ so, given our choice of simple roots (see appendix \ref{sec: int octs}), the generator matrix is,

\begin{equation}
    M_{\text{gen}} = \begin{pmatrix}
        0 & 0 & 0 & 0 & 0 & 0 & 0 & \frac{1}{2}\\ 
0 & 0 & 0 & 0 & 0 & -\frac{1}{2} & 1 & -\frac{1}{2}\\ 
-\frac{1}{2} & -\frac{1}{2} & \frac{1}{2} & \frac{1}{2} & -\frac{1}{2} & 0 & 0 & 0\\ 
\frac{1}{2} & -\frac{1}{2} & -\frac{1}{2} & \frac{1}{2} & -\frac{1}{2} & \frac{1}{2} & 0 & 0\\ 
-\frac{1}{2} & 0 & 0 & 0 & 0 & \frac{1}{2} & 0 & -\frac{1}{2}\\ 
-\frac{1}{2} & 0 & 0 & 0 & 0 & 0 & 0 & \frac{1}{2}\\ 
0 & \frac{1}{2} & -\frac{1}{2} & \frac{1}{2} & -\frac{1}{2} & 0 & 0 & 0\\ 
0 & -\frac{1}{2} & -\frac{1}{2} & \frac{1}{2} & \frac{1}{2} & -\frac{1}{2} & 0 & 0 
    \end{pmatrix}
\end{equation}
Take one of the generators $g^{(i)}_{a} = L_{e_{b}}\circ L_{e_{c}}$ of one of the groups $G_{i}$. To find its matrix representation, denoted $T_{a}^{(i)}$, one must solve the following equations,

\begin{equation}
    T_{a}^{(i)}\;M_{\text{gen}} = g_{a}^{(i)}(M_{gen}) \quad \Rightarrow\quad  T_{a}^{(i)} = g_{a}^{(i)}(M_{gen})\;M_{\text{gen}}^{-1}\;
\end{equation}
where by $g_{a}^{(i)}(M_{gen})$, we mean that we take each column of $M_{\text{gen}}$ and treat it as an octonion, then we act on it by $g_{a}^{(i)}$ and rewrite it as a column vector. Since, by construction, acting with $g_{a}^{(i)}$ on the lattice preserves it (as the integral octonions are closed under left multiplication) the corresponding matrix $T_{a}^{(i)}$ will necessarily be orthogonal and an automorphism of the lattice. In other words it satisfies the equation \cite{conway2013sphere} (see appendix \ref{sec:class trans}),
\begin{equation}
    M_{\text{gen}}\;C = T_{a}^{(i)}\;M_{\text{gen}}
\end{equation}
where $C\in \op{GL}(8,\mathbb{Z})$. 

Also by construction, these matrices preserve a hyperK\"ahler structure on $\mathbb{R}^{8}$. We prove this in appendix $\ref{sec: basis change mats}$ where an explicit basis change is given for each matrix group such that they preserve the standard hyperK\"ahler structure on $\mathbb{R}^{8}$. 

In the following we display for each $G_{i}$ the generators in terms of left multiplication by octonions and the corresponding matrices acting on $\mathbb{R}^{8}$.

\vspace{2em}
\noindent \underline{(I) $\;\; G_{1} = \pi_{2}(\pi_{1}^{-1}(W(D_{4})))$}
\begin{equation}
    g_{1}^{(1)} = L_{e_{7}}\circ L_{e_{2}},\quad g_{2}^{(1)} = L_{e_{7}}\circ L_{e_{3}},\quad g_{3}^{(1)} = L_{e_{7}}\circ L_{e_{4}},\quad g_{4}^{(1)} = L_{e_{7}}\circ L_{e_{5}}
\end{equation}

\begin{align*}
    T^{(1)}_{1} &= 
 \begin{psmallmatrix}
 0 & 0 & -\frac{1}{2} & \frac{1}{2} & 0 & 0 & \frac{1}{2} & \frac{1}{2} \\
 0 & 0 & \frac{1}{2} & \frac{1}{2} & 0 & 0 & -\frac{1}{2} & \frac{1}{2} \\
 \frac{1}{2} & -\frac{1}{2} & 0 & 0 & -\frac{1}{2} & -\frac{1}{2} & 0 & 0 \\
 -\frac{1}{2} & -\frac{1}{2} & 0 & 0 & -\frac{1}{2} & \frac{1}{2} & 0 & 0 \\
 0 & 0 & \frac{1}{2} & \frac{1}{2} & 0 & 0 & \frac{1}{2} & -\frac{1}{2} \\
 0 & 0 & \frac{1}{2} & -\frac{1}{2} & 0 & 0 & \frac{1}{2} & \frac{1}{2} \\
 -\frac{1}{2} & \frac{1}{2} & 0 & 0 & -\frac{1}{2} & -\frac{1}{2} & 0 & 0 \\
 -\frac{1}{2} & -\frac{1}{2} & 0 & 0 & \frac{1}{2} & -\frac{1}{2} & 0 & 0
\end{psmallmatrix},\quad T^{(1)}_{2} = \begin{psmallmatrix}
    0 & 0 & -\frac{1}{2} & -\frac{1}{2} & 0 & 0 & \frac{1}{2} & -\frac{1}{2} \\
 0 & 0 & -\frac{1}{2} & \frac{1}{2} & 0 & 0 & \frac{1}{2} & \frac{1}{2} \\
 \frac{1}{2} & \frac{1}{2} & 0 & 0 & -\frac{1}{2} & \frac{1}{2} & 0 & 0 \\
 \frac{1}{2} & -\frac{1}{2} & 0 & 0 & \frac{1}{2} & \frac{1}{2} & 0 & 0 \\
 0 & 0 & \frac{1}{2} & -\frac{1}{2} & 0 & 0 & \frac{1}{2} & \frac{1}{2} \\
 0 & 0 & -\frac{1}{2} & -\frac{1}{2} & 0 & 0 & -\frac{1}{2} & \frac{1}{2} \\
 -\frac{1}{2} & -\frac{1}{2} & 0 & 0 & -\frac{1}{2} & \frac{1}{2} & 0 & 0 \\
 \frac{1}{2} & -\frac{1}{2} & 0 & 0 & -\frac{1}{2} & -\frac{1}{2} & 0 & 0
\end{psmallmatrix}\\
\nonumber \\
T^{(1)}_{3} &= \begin{psmallmatrix}
     0 & 0 & \frac{1}{2} & -\frac{1}{2} & 0 & 0 & -\frac{1}{2} & \frac{1}{2} \\
 0 & 0 & -\frac{1}{2} & -\frac{1}{2} & 0 & 0 & -\frac{1}{2} & -\frac{1}{2} \\
 -\frac{1}{2} & \frac{1}{2} & 0 & 0 & \frac{1}{2} & -\frac{1}{2} & 0 & 0 \\
 \frac{1}{2} & \frac{1}{2} & 0 & 0 & -\frac{1}{2} & -\frac{1}{2} & 0 & 0 \\
 0 & 0 & -\frac{1}{2} & \frac{1}{2} & 0 & 0 & -\frac{1}{2} & \frac{1}{2} \\
 0 & 0 & \frac{1}{2} & \frac{1}{2} & 0 & 0 & -\frac{1}{2} & -\frac{1}{2} \\
 \frac{1}{2} & \frac{1}{2} & 0 & 0 & \frac{1}{2} & \frac{1}{2} & 0 & 0 \\
 -\frac{1}{2} & \frac{1}{2} & 0 & 0 & -\frac{1}{2} & \frac{1}{2} & 0 & 0
\end{psmallmatrix},\quad T^{(1)}_{4} = \begin{psmallmatrix}
     0 & 0 & -\frac{1}{2} & \frac{1}{2} & 0 & 0 & -\frac{1}{2} & -\frac{1}{2} \\
 0 & 0 & \frac{1}{2} & \frac{1}{2} & 0 & 0 & \frac{1}{2} & -\frac{1}{2} \\
 \frac{1}{2} & -\frac{1}{2} & 0 & 0 & \frac{1}{2} & \frac{1}{2} & 0 & 0 \\
 -\frac{1}{2} & -\frac{1}{2} & 0 & 0 & \frac{1}{2} & -\frac{1}{2} & 0 & 0 \\
 0 & 0 & -\frac{1}{2} & -\frac{1}{2} & 0 & 0 & \frac{1}{2} & -\frac{1}{2} \\
 0 & 0 & -\frac{1}{2} & \frac{1}{2} & 0 & 0 & \frac{1}{2} & \frac{1}{2} \\
 \frac{1}{2} & -\frac{1}{2} & 0 & 0 & -\frac{1}{2} & -\frac{1}{2} & 0 & 0 \\
 \frac{1}{2} & \frac{1}{2} & 0 & 0 & \frac{1}{2} & -\frac{1}{2} & 0 & 0
\end{psmallmatrix}\nonumber
\end{align*}

\vspace{2em}

\noindent\underline{(II) $\;\; G_{2} = \pi_{2}(\pi_{1}^{-1}(W^{+}(A_{5})))$}
\begin{equation}
    g_{1}^{(2)} = L_{e_{2}}\circ L_{e_{4}},\quad g_{2}^{(2)} = L_{e_{4}}\circ L_{e_{5}},\quad g_{3}^{(2)} = L_{e_{5}}\circ L_{e_{6}},\quad g_{4}^{(2)} = L_{e_{6}}\circ L_{e_{7}}
\end{equation}

\begin{align}
    T^{(2)}_{1} &= \begin{psmallmatrix}
      \frac{1}{2} & \frac{1}{2} & 0 & 0 & -\frac{1}{2} & \frac{1}{2} & 0 & 0 \\
 -\frac{1}{2} & \frac{1}{2} & 0 & 0 & -\frac{1}{2} & -\frac{1}{2} & 0 & 0 \\
 0 & 0 & \frac{1}{2} & -\frac{1}{2} & 0 & 0 & \frac{1}{2} & \frac{1}{2} \\
 0 & 0 & \frac{1}{2} & \frac{1}{2} & 0 & 0 & \frac{1}{2} & -\frac{1}{2} \\
 \frac{1}{2} & \frac{1}{2} & 0 & 0 & \frac{1}{2} & -\frac{1}{2} & 0 & 0 \\
 -\frac{1}{2} & \frac{1}{2} & 0 & 0 & \frac{1}{2} & \frac{1}{2} & 0 & 0 \\
 0 & 0 & -\frac{1}{2} & -\frac{1}{2} & 0 & 0 & \frac{1}{2} & -\frac{1}{2} \\
 0 & 0 & -\frac{1}{2} & \frac{1}{2} & 0 & 0 & \frac{1}{2} & \frac{1}{2}
    \end{psmallmatrix},\quad T^{(2)}_{2} = \begin{psmallmatrix}
        \frac{1}{2} & \frac{1}{2} & 0 & 0 & \frac{1}{2} & \frac{1}{2} & 0 & 0 \\
 -\frac{1}{2} & \frac{1}{2} & 0 & 0 & -\frac{1}{2} & \frac{1}{2} & 0 & 0 \\
 0 & 0 & \frac{1}{2} & -\frac{1}{2} & 0 & 0 & \frac{1}{2} & -\frac{1}{2} \\
 0 & 0 & \frac{1}{2} & \frac{1}{2} & 0 & 0 & -\frac{1}{2} & -\frac{1}{2} \\
 -\frac{1}{2} & \frac{1}{2} & 0 & 0 & \frac{1}{2} & -\frac{1}{2} & 0 & 0 \\
 -\frac{1}{2} & -\frac{1}{2} & 0 & 0 & \frac{1}{2} & \frac{1}{2} & 0 & 0 \\
 0 & 0 & -\frac{1}{2} & \frac{1}{2} & 0 & 0 & \frac{1}{2} & -\frac{1}{2} \\
 0 & 0 & \frac{1}{2} & \frac{1}{2} & 0 & 0 & \frac{1}{2} & \frac{1}{2}
    \end{psmallmatrix}\nonumber\\
    \nonumber\\
    T^{(2)}_{3} &= \begin{psmallmatrix}
        \frac{1}{2} & \frac{1}{2} & -\frac{1}{2} & \frac{1}{2} & 0 & 0 & 0 & 0 \\
 -\frac{1}{2} & \frac{1}{2} & 0 & 0 & 0 & 0 & \frac{1}{2} & -\frac{1}{2} \\
 \frac{1}{2} & 0 & \frac{1}{2} & 0 & \frac{1}{2} & 0 & 0 & -\frac{1}{2} \\
 -\frac{1}{2} & 0 & 0 & \frac{1}{2} & \frac{1}{2} & 0 & -\frac{1}{2} & 0 \\
 0 & 0 & -\frac{1}{2} & -\frac{1}{2} & \frac{1}{2} & -\frac{1}{2} & 0 & 0 \\
 0 & 0 & 0 & 0 & \frac{1}{2} & \frac{1}{2} & \frac{1}{2} & \frac{1}{2} \\
 0 & -\frac{1}{2} & 0 & \frac{1}{2} & 0 & -\frac{1}{2} & \frac{1}{2} & 0 \\
 0 & \frac{1}{2} & \frac{1}{2} & 0 & 0 & -\frac{1}{2} & 0 & \frac{1}{2}
    \end{psmallmatrix},\quad T^{(2)}_{4} = \begin{psmallmatrix}
       \frac{1}{2} & 0 & -\frac{1}{2} & 0 & 0 & \frac{1}{2} & -\frac{1}{2} & 0 \\
 0 & \frac{1}{2} & 0 & \frac{1}{2} & \frac{1}{2} & 0 & 0 & -\frac{1}{2} \\
 \frac{1}{2} & 0 & \frac{1}{2} & 0 & \frac{1}{2} & 0 & 0 & \frac{1}{2} \\
 0 & -\frac{1}{2} & 0 & \frac{1}{2} & 0 & -\frac{1}{2} & -\frac{1}{2} & 0 \\
 0 & -\frac{1}{2} & -\frac{1}{2} & 0 & \frac{1}{2} & 0 & \frac{1}{2} & 0 \\
 -\frac{1}{2} & 0 & 0 & \frac{1}{2} & 0 & \frac{1}{2} & 0 & \frac{1}{2} \\
 \frac{1}{2} & 0 & 0 & \frac{1}{2} & -\frac{1}{2} & 0 & \frac{1}{2} & 0 \\
 0 & \frac{1}{2} & -\frac{1}{2} & 0 & 0 & -\frac{1}{2} & 0 & \frac{1}{2}
    \end{psmallmatrix}\nonumber
\end{align}

\vspace{2em}

\noindent\underline{(III) $\;\;G_{3} = \pi_{2}(\pi_{1}^{-1}(W^{+}(D_{5})))$}

\begin{equation}
    g_{1}^{(3)} = L_{e_{1}}\circ L_{e_{3}},\quad g_{2}^{(3)} = L_{e_{3}}\circ L_{e_{4}},\quad g_{3}^{(3)} = L_{e_{2}}\circ L_{e_{4}},\quad g_{4}^{(3)} = L_{e_{5}}\circ L_{e_{4}}
\end{equation}

\begin{align}
    T^{(3)}_{1} &= 
 \begin{psmallmatrix}
 \frac{1}{2} & 0 & -\frac{1}{2} & 0 & 0 & \frac{1}{2} & -\frac{1}{2} & 0 \\
 0 & \frac{1}{2} & 0 & -\frac{1}{2} & -\frac{1}{2} & 0 & 0 & \frac{1}{2} \\
 \frac{1}{2} & 0 & \frac{1}{2} & 0 & 0 & -\frac{1}{2} & -\frac{1}{2} & 0 \\
 0 & \frac{1}{2} & 0 & \frac{1}{2} & \frac{1}{2} & 0 & 0 & \frac{1}{2} \\
 0 & \frac{1}{2} & 0 & -\frac{1}{2} & \frac{1}{2} & 0 & 0 & -\frac{1}{2} \\
 -\frac{1}{2} & 0 & \frac{1}{2} & 0 & 0 & \frac{1}{2} & -\frac{1}{2} & 0 \\
 \frac{1}{2} & 0 & \frac{1}{2} & 0 & 0 & \frac{1}{2} & \frac{1}{2} & 0 \\
 0 & -\frac{1}{2} & 0 & -\frac{1}{2} & \frac{1}{2} & 0 & 0 & \frac{1}{2}
\end{psmallmatrix},\quad T^{(3)}_{2} = \begin{psmallmatrix}
    \frac{1}{2} & -\frac{1}{2} & 0 & 0 & \frac{1}{2} & \frac{1}{2} & 0 & 0 \\
 \frac{1}{2} & \frac{1}{2} & 0 & 0 & -\frac{1}{2} & \frac{1}{2} & 0 & 0 \\
 0 & 0 & \frac{1}{2} & -\frac{1}{2} & 0 & 0 & -\frac{1}{2} & -\frac{1}{2} \\
 0 & 0 & \frac{1}{2} & \frac{1}{2} & 0 & 0 & -\frac{1}{2} & \frac{1}{2} \\
 -\frac{1}{2} & \frac{1}{2} & 0 & 0 & \frac{1}{2} & \frac{1}{2} & 0 & 0 \\
 -\frac{1}{2} & -\frac{1}{2} & 0 & 0 & -\frac{1}{2} & \frac{1}{2} & 0 & 0 \\
 0 & 0 & \frac{1}{2} & \frac{1}{2} & 0 & 0 & \frac{1}{2} & -\frac{1}{2} \\
 0 & 0 & \frac{1}{2} & -\frac{1}{2} & 0 & 0 & \frac{1}{2} & \frac{1}{2}
\end{psmallmatrix}\\
\nonumber \\
T^{(3)}_{3} &= \begin{psmallmatrix}
     \frac{1}{2} & \frac{1}{2} & 0 & 0 & -\frac{1}{2} & \frac{1}{2} & 0 & 0 \\
 -\frac{1}{2} & \frac{1}{2} & 0 & 0 & -\frac{1}{2} & -\frac{1}{2} & 0 & 0 \\
 0 & 0 & \frac{1}{2} & -\frac{1}{2} & 0 & 0 & \frac{1}{2} & \frac{1}{2} \\
 0 & 0 & \frac{1}{2} & \frac{1}{2} & 0 & 0 & \frac{1}{2} & -\frac{1}{2} \\
 \frac{1}{2} & \frac{1}{2} & 0 & 0 & \frac{1}{2} & -\frac{1}{2} & 0 & 0 \\
 -\frac{1}{2} & \frac{1}{2} & 0 & 0 & \frac{1}{2} & \frac{1}{2} & 0 & 0 \\
 0 & 0 & -\frac{1}{2} & -\frac{1}{2} & 0 & 0 & \frac{1}{2} & -\frac{1}{2} \\
 0 & 0 & -\frac{1}{2} & \frac{1}{2} & 0 & 0 & \frac{1}{2} & \frac{1}{2}
\end{psmallmatrix},\quad T^{(3)}_{4} = \begin{psmallmatrix}
     \frac{1}{2} & -\frac{1}{2} & 0 & 0 & -\frac{1}{2} & -\frac{1}{2} & 0 & 0 \\
 \frac{1}{2} & \frac{1}{2} & 0 & 0 & \frac{1}{2} & -\frac{1}{2} & 0 & 0 \\
 0 & 0 & \frac{1}{2} & \frac{1}{2} & 0 & 0 & -\frac{1}{2} & \frac{1}{2} \\
 0 & 0 & -\frac{1}{2} & \frac{1}{2} & 0 & 0 & \frac{1}{2} & \frac{1}{2} \\
 \frac{1}{2} & -\frac{1}{2} & 0 & 0 & \frac{1}{2} & \frac{1}{2} & 0 & 0 \\
 \frac{1}{2} & \frac{1}{2} & 0 & 0 & -\frac{1}{2} & \frac{1}{2} & 0 & 0 \\
 0 & 0 & \frac{1}{2} & -\frac{1}{2} & 0 & 0 & \frac{1}{2} & \frac{1}{2} \\
 0 & 0 & -\frac{1}{2} & -\frac{1}{2} & 0 & 0 & -\frac{1}{2} & \frac{1}{2}
\end{psmallmatrix}\nonumber
\end{align}

\vspace{2em}

Now that we have the three matrix groups $G_{i} = \langle T_{1}^{(i)},T_{2}^{(i)},T_{3}^{(i)},T_{4}^{(i)}\rangle$, one can use GAP \cite{GAP4} to efficiently produce a list of all the subgroups of each $G_{i}$, up to conjugacy. Groups that are conjugate inside $G_{i}$ will give identical singular sets (this is a result of the fact that $G_{i}$ is a subgroup of the automorphism group of the lattice). It is still possible that some of these groups are conjugate under $\op{GL}(8,\mathbb{Z})$ when written in the lattice basis and therefore give rise to equivalent space groups (see appendix \ref{sec:class trans}) but this is not important for the results of this paper. In the next section we list the resulting lists of subgroups.

\subsection{List of model hyperK\"ahler orbifolds}\label{sec: list of subgroups}

We include below the list of subgroups for each of the $G_{i}$. We omit any Abelian groups and also non-Abelian groups of the form $\Gamma\times \Gamma'$ where $\Gamma, \Gamma' = \{\mathbb{1}\}, \mathbb{Z}_{2},\mathbb{Z}_{3},\mathbb{Z}_{4},\mathbb{Z}_{6},\mathbb{D}_{2}, \mathbb{D}_{3}, \mathbb{T}$ acting on $T^{8}$ as $(T^{4}/\Gamma)\times (T^{4}/\Gamma')$ since these are known to admit a resolution by $K3\times K3$ or $K3\times T^{4}$. In the table we include the ID of the group in GAP's Small Group library and a list of its generators. The generators are taken straight from the output of the GAP code; in the $i$-th table $F_{j}$ corresponds to $T_{j}^{(i)}$ from \ref{sec: mat reps}. Those rows that are highlighted are groups from the list in section \ref{sec: symp resolutions} that are known to admit a symplectic resolution and are acting in the representation stated there. Groups for which it is still open as to whether they admit a symplectic resolution are marked by stars. Using these tables together with the matrices and the lattice in \ref{sec: mat reps}, one can reproduce all the orbifolds that we consider.

The second column in each table provides a name for each group. These names are far from unique but are included to give some insight into the structure of each group. The notation we use is the following,
\begin{itemize}
    \item $S_{n}$ denotes the symmetric group on $n$ letters.
    \item $Q_{8}$ denotes the quaternion group.
    \item $\mathbb{Z}_{n}$ denotes the cyclic group of order $n$.
    \item $D_{n}$ denotes the dihedral group of order $2n$.
    \item $\mathbb{D}_{n}$ denotes the binary dihedral group of order $4n$.
    \item $\mathbb{T}$, $\mathbb{O}$ and $\mathbb{I}$ denote the binary tetrahedral, binary octahedral and binary icosahedral groups respectively.
    \item $M_{n}(2)$ denotes the modular maximal cyclic group of order $2^{n}$.
    \item $QD_{n}$ denotes the quasi-dihedral group of order $2^{n}$.
    \item $\ltimes$ and 
    $\rtimes$ denote semidirect products, $N\cdot H$ denotes a non-split extension of $H$ by $N$, $H\wr S_{n}$ denotes the wreath product $H^{n}\rtimes S_{n}$ and $\circ$ denotes a central product.
\end{itemize}

\newpage
\begin{scriptsize}
\begin{longtable}{|l|l|l|}
\hline
\multicolumn{3}{|c|}{}\\
\multicolumn{3}{|c|}{\large{\textbf{Subgroups of $G_{1}$}}}\\
\multicolumn{3}{|c|}{}\\
\hline
& &\\
Group ID & Name & Generators \label{tab:table 1} \\
& &\\
\hline
& &\\
\rowcolor{LightSteelBlue}\textbf{1:} \hspace{1em}$ [6,1] $ & $S_{3}$ & $ [F_{2} F_{3} F_{2} F_{1}^{-1},F_{3} F_{4} F_{3} F_{1} F_{2}^{-1} F_{3}^{-1} F_{4}^{-1}] $ \\ 
 &&\\ 
\textbf{2:} \hspace{1em}$ [8,4] $ & $Q_{8}$ & $ [F_{1}^{-2},F_{3} F_{4} F_{2}^{-1} F_{3}^{-1},F_{1} F_{3} F_{4} F_{2}^{-1} F_{3}^{-1} F_{1}^{-1}] $ \\ 
 &&\\ 
\textbf{3:} \hspace{1em}$ [8,4] $ & $Q_{8}$ & $ [F_{1}^{-2},F_{3} F_{2} F_{1}^{-1} F_{3}^{-1},F_{4} F_{3} F_{2} F_{1}^{-1} F_{3}^{-1} F_{4}^{-1}] $ \\ 
 &&\\ 
\textbf{4:} \hspace{1em}$ [8,4] $ & $Q_{8}$ & $ [F_{1}^{-2},F_{1} F_{3} F_{4} F_{1}^{-1} F_{3}^{-1} F_{1}^{-1},F_{1} F_{2} F_{3} F_{4} F_{1}^{-1} F_{3}^{-1} F_{2}^{-1} F_{1}^{-1}] $ \\ 
 &&\\ 
\textbf{5:} \hspace{1em}$ [8,4] $ & $Q_{8}$ & $ [F_{3}^{-1} F_{2}^{2} F_{3}^{-1},F_{3}^{-1} F_{2}^{-2} F_{3}^{-1},F_{2} F_{3} F_{2} F_{4}^{-1} F_{3}^{-1} F_{1}^{-1},F_{3} F_{1} F_{2} F_{3}^{-1} F_{4}^{-1} F_{2}^{-1}] $ \\ 
 &&\\ 
\rowcolor{LightSteelBlue}\textbf{6:} \hspace{1em}$ [8,3] $ & $D_{4}$ & $ [F_{1}^{-2},F_{2} F_{1}^{-1},F_{3} F_{1} F_{4} F_{3} F_{2}^{-2} F_{3}^{-1} F_{4}^{-1} F_{1}^{-1} F_{3}^{-1},F_{1} F_{3} F_{1} F_{2} F_{3} F_{4} F_{3}^{-1} F_{2}^{-1} F_{1}^{-1} F_{3}^{-1} F_{4}^{-1}] $ \\ 
 &&\\ 
\rowcolor{LightSteelBlue}\textbf{7:} \hspace{1em}$ [8,3] $ & $D_{4}$ & $ [F_{1}^{-2},F_{2}^{-2},F_{1} F_{2} F_{4}^{-1},F_{3} F_{1} F_{2} F_{3} F_{4} F_{3}^{-1} F_{2}^{-1} F_{1}^{-1} F_{3}^{-1} F_{1}^{-1}] $ \\ 
 &&\\ 
\rowcolor{LightSteelBlue}\textbf{8:} \hspace{1em}$ [8,3] $ & $D_{4}$ & $ [F_{1}^{-2},F_{4} F_{1}^{-1},F_{3} F_{1} F_{4} F_{3} F_{2}^{-2} F_{3}^{-1} F_{4}^{-1} F_{1}^{-1} F_{3}^{-1},(F_{1} F_{2} F_{3})^{2} F_{4}^{-1} F_{3}^{-1} F_{2}^{-1} F_{1}^{-1} F_{3}^{-1}] $ \\ 
 &&\\ 
\textbf{9:} \hspace{1em}$ [8,4] $ & $Q_{8}$ & $ [F_{1}^{-2},F_{2} F_{3} F_{2}^{-1},F_{3} F_{4} F_{3}^{-1}] $ \\ 
 &&\\ 
\textbf{10:} \hspace{1em}$ [8,4] $ & $Q_{8}$ & $ [F_{1}^{-2},F_{2} F_{3} F_{2}^{-1},F_{1} F_{2} F_{3} F_{4} F_{3}^{-1} F_{2}^{-1} F_{1}^{-1}] $ \\ 
 &&\\ 
\textbf{11:} \hspace{1em}$ [8,4] $ & $Q_{8}$ & $ [F_{1}^{-2},F_{1} F_{3} F_{1}^{-1},F_{2} F_{3} F_{2}^{-1}] $ \\ 
 &&\\ 
\rowcolor{LightSteelBlue}\textbf{12:} \hspace{1em}$ [8,3] $ & $D_{4}$ & $ [F_{1},F_{3} F_{1} F_{4} F_{3} F_{2}^{-2} F_{3}^{-1} F_{4}^{-1} F_{1}^{-1} F_{3}^{-1},F_{2} F_{3} F_{1} F_{2} F_{3} F_{4} F_{3}^{-1} F_{2}^{-1} F_{1}^{-1} F_{3}^{-1} F_{4}^{-1}] $ \\ 
 &&\\ 
\textbf{13:} \hspace{1em}$ [8,4] $ & $Q_{8}$ & $ [F_{1}^{-2},F_{4} F_{2}^{-1},F_{3} F_{1} F_{2} F_{3} F_{4} F_{3}^{-1} F_{2}^{-1} F_{1}^{-1} F_{3}^{-1} F_{2}^{-1}] $ \\ 
 &&\\ 
\textbf{14:} \hspace{1em}$ [12,1] $ & $\mathbb{D}_{3}$ & $ [F_{1},F_{2} F_{3} F_{2}^{-1}] $ \\ 
 &&\\ 
\textbf{15:} \hspace{1em}$ [12,4] $ & $D_{6}$ & $ [F_{1}^{-2},F_{2} F_{3} F_{2}^{-1} F_{1}^{-1},F_{3} F_{4} F_{3} F_{1} F_{2}^{-1} F_{3}^{-1} F_{4}^{-1}] $ \\ 
 &&\\ 
\textbf{16:} \hspace{1em}$ [16,12] $ & $\mathbb{Z}_{2}\times Q_{8}$ & $ [F_{1}^{-2},F_{4} F_{2}^{-1},F_{3} F_{4} F_{2}^{-1} F_{3}^{-1},F_{1} F_{3} F_{4} F_{2}^{-1} F_{3}^{-1} F_{1}^{-1}] $ \\ 
 &&\\ 
\textbf{17:} \hspace{1em}$ [16,12] $ & $\mathbb{Z}_{2}\times Q_{8}$ & $ [F_{1}^{-2},F_{2} F_{1}^{-1},F_{3} F_{2} F_{1}^{-1} F_{3}^{-1},F_{4} F_{3} F_{2} F_{1}^{-1} F_{3}^{-1} F_{4}^{-1}] $ \\ 
 &&\\ 
\textbf{18:} \hspace{1em}$ [16,12] $ & $\mathbb{Z}_{2}\times Q_{8}$ & $ [F_{1}^{-2},F_{4} F_{1}^{-1},F_{3} F_{4} F_{1}^{-1} F_{3}^{-1},F_{2} F_{3} F_{4} F_{1}^{-1} F_{3}^{-1} F_{2}^{-1}] $ \\ 
 &&\\ 
\textbf{19:} \hspace{1em}$ [16,12] $ & $\mathbb{Z}_{2}\times Q_{8}$ & $ [F_{1}^{-2},F_{2} F_{1}^{-1},F_{4} F_{1}^{-1},F_{3} F_{1} F_{2} F_{3} F_{4} F_{3}^{-1} F_{2}^{-1} F_{1}^{-1} F_{3}^{-1} F_{1}^{-1}] $ \\ 
 &&\\ 
\textbf{20:} \hspace{1em}$ [16,6] $ & $M_{4}(2)$ & $ [F_{1}^{-2},F_{2} F_{4} F_{3}^{-1},F_{1} F_{3} F_{4} F_{2}^{-1} F_{3}^{-1} F_{1}^{-1}] $ \\ 
 &&\\ 
\textbf{21:} \hspace{1em}$ [16,4] $ & $\mathbb{Z}_{4}\rtimes \mathbb{Z}_{4}$ & $ [F_{1}^{-2},F_{3} F_{4} F_{1}^{-1} F_{3}^{-1},F_{1} F_{3} F_{4} F_{2}^{-1} F_{3}^{-1} F_{1}^{-1}] $ \\ 
 &&\\ 
\textbf{22:} \hspace{1em}$ [16,4] $ & $\mathbb{Z}_{4}\rtimes \mathbb{Z}_{4}$ & $ [F_{1}^{-2},F_{4} F_{1}^{-1},F_{3} F_{2}^{-2} F_{3}^{-1},F_{1} F_{3} F_{2} F_{1}^{-1} F_{3}^{-1} F_{1}^{-1}] $ \\ 
 &&\\ 
\textbf{23:} \hspace{1em}$ [16,4] $ & $\mathbb{Z}_{4}\rtimes \mathbb{Z}_{4}$ & $ [F_{1}^{-2},F_{4} F_{2}^{-1},F_{3} F_{2} F_{1}^{-1} F_{3}^{-1}] $ \\ 
 &&\\ 
\textbf{24:} \hspace{1em}$ [16,13] $ & $\mathbb{Z}_{4}\circ D_{4}$ & $ [F_{1},F_{2},F_{3} F_{1} F_{4} F_{3} F_{2} F_{1}^{-1} F_{3}^{-1} F_{4}^{-1} F_{1}^{-1} F_{3}^{-1}] $ \\ 
 &&\\ 
\textbf{25:} \hspace{1em}$ [16,13] $ & $\mathbb{Z}_{4}\circ D_{4}$ & $ [F_{1},F_{4} F_{2}^{-1},F_{3} F_{1} F_{2} F_{3} F_{4} F_{3}^{-1} F_{2}^{-1} F_{1}^{-1} F_{3}^{-1}] $ \\ 
 &&\\ 
\textbf{26:} \hspace{1em}$ [16,13] $ & $\mathbb{Z}_{4}\circ D_{4}$ & $ [F_{1},F_{4},F_{3} F_{1} F_{2} F_{3} F_{4} F_{3}^{-1} F_{2}^{-1} F_{1}^{-1} F_{3}^{-1} F_{2}^{-1}] $ \\ 
 &&\\ 
\textbf{27:} \hspace{1em}$ [16,6] $ & $M_{4}(2)$ & $ [F_{1}^{-2},F_{2} F_{1}^{-1},F_{3} F_{1} F_{2} F_{3}^{-1} F_{4}^{-1}] $ \\ 
 &&\\ 
\textbf{28:} \hspace{1em}$ [16,6] $ & $M_{4}(2)$ & $ [F_{1}^{-2},F_{1} F_{3} F_{4}^{-1},F_{1} F_{2} F_{3} F_{4} F_{1}^{-1} F_{3}^{-1} F_{2}^{-1} F_{1}^{-1}] $ \\ 
 &&\\ 
\textbf{29:} \hspace{1em}$ [16,8] $ & $QD_{16}$ & $ [F_{1}^{-2},F_{2} F_{1}^{-1},F_{3} F_{2}^{-2} F_{3}^{-1},F_{1} F_{3} F_{1} F_{2}^{-1} F_{3}^{-1},F_{4} F_{3} F_{2} F_{1}^{-1} F_{3}^{-1} F_{4}^{-1}] $ \\ 
 &&\\ 
\textbf{30:} \hspace{1em}$ [16,13] $ & $\mathbb{Z}_{4}\circ D_{4}$ & $ [F_{1}^{-2},F_{2} F_{1}^{-1},F_{4},F_{3} F_{1} F_{2} F_{3} F_{4} F_{3}^{-1} F_{2}^{-1} F_{1}^{-1} F_{3}^{-1} F_{1}^{-1}] $ \\ 
 &&\\ 
\textbf{31:} \hspace{1em}$ [16,8] $ & $QD_{16}$ & $ [F_{1}^{-2},F_{2}^{-2},F_{1} F_{2} F_{4}^{-1},F_{3} F_{4} F_{2}^{-1} F_{3}^{-1}] $ \\ 
 &&\\ 
\textbf{32:} \hspace{1em}$ [16,12] $ & $\mathbb{Z}_{2}\times Q_{8}$ & $ [F_{1}^{-2},F_{2},F_{4},F_{3} F_{1} F_{2} F_{3} F_{4} F_{3}^{-1} F_{2}^{-1} F_{1}^{-1} F_{3}^{-1}] $ \\ 
 &&\\ 
\textbf{33:} \hspace{1em}$ [16,8] $ & $QD_{16}$ & $ [F_{1}^{-2},F_{4} F_{1}^{-1},F_{1} F_{3} F_{1} F_{4}^{-1} F_{3}^{-1},F_{2} F_{3} F_{4} F_{1}^{-1} F_{3}^{-1} F_{2}^{-1}] $ \\ 
 &&\\ 
\textbf{34:} \hspace{1em}$ [16,9] $ & $\mathbb{D}_{4}$ & $ [F_{1}^{-2},F_{2} F_{3} F_{2}^{-1},F_{3} F_{4} F_{3}^{-1},F_{1} F_{3} F_{4} F_{2}^{-1} F_{3}^{-1} F_{1}^{-1}] $ \\ 
 &&\\ 
\textbf{35:} \hspace{1em}$ [16,9] $ & $\mathbb{D}_{4}$ & $ [F_{1}^{-2},F_{2} F_{3} F_{2}^{-1},F_{1} F_{3} F_{4} F_{1}^{-1} F_{3}^{-1} F_{1}^{-1}] $ \\ 
 &&\\ 
\textbf{36:} \hspace{1em}$ [16,9] $ & $\mathbb{D}_{4}$ & $ [F_{1}^{-2},F_{1} F_{3} F_{1}^{-1},F_{2} F_{3} F_{2}^{-1},F_{4} F_{3} F_{2} F_{1}^{-1} F_{3}^{-1} F_{4}^{-1}] $ \\ 
 &&\\ 
\textbf{37:} \hspace{1em}$ [24,3] $ & $\mathbb{T}$ & $ [F_{1}^{-2},F_{1} F_{4} F_{3}^{-1} F_{1}^{-1},F_{2} F_{4} F_{3}^{-1} F_{2}^{-1}] $ \\ 
 &&\\ 
\textbf{38:} \hspace{1em}$ [24,3] $ & $\mathbb{T}$ & $ [F_{1}^{-2},F_{3} F_{2}^{-1},F_{1} F_{3} F_{4} F_{3}^{-1} F_{2}^{-1} F_{1}^{-1}] $ \\ 
 &&\\ 
\textbf{39:} \hspace{1em}$ [24,3] $ & $\mathbb{T}$ & $ [F_{1}^{-2},F_{2} F_{3} F_{2}^{-1} F_{1}^{-1},F_{3} F_{4} F_{3}^{-1} F_{1}^{-1}] $ \\ 
 &&\\ 
\textbf{40:} \hspace{1em}$ [24,3] $ & $\mathbb{T}$ & $ [F_{1} F_{4} F_{3}^{-1} F_{1}^{-1},F_{2} F_{3} F_{2} F_{1}^{-1}] $ \\ 
 &&\\ 
\textbf{41:} \hspace{1em}$ [24,8] $ & $\mathbb{Z}_{3}\rtimes D_{4}$ & $ [F_{1},F_{2} F_{3} F_{2}^{-1},F_{3} F_{4} F_{3} F_{1} F_{2}^{-1} F_{3}^{-1} F_{4}^{-1}] $ \\ 
 &&\\ 
\textbf{42:} \hspace{1em}$ [32,8] $ & $\mathbb{Z}_{4}\cdot D_{4}$ & $ [F_{1}^{-2},F_{2} F_{1}^{-1},F_{4} F_{1}^{-1},F_{1} F_{3} F_{1} F_{4}^{-1} F_{3}^{-1}] $ \\ 
 &&\\ 
\textbf{43:} \hspace{1em}$ [32,8] $ & $\mathbb{Z}_{4}\cdot D_{4}$ & $ [F_{1}^{-2},F_{2} F_{1}^{-1},F_{4} F_{1}^{-1},F_{3} F_{2}^{-2} F_{3}^{-1},F_{1} F_{3} F_{1} F_{2}^{-1} F_{3}^{-1}] $ \\ 
 &&\\ 
\rowcolor{LightSteelBlue}\textbf{44:} \hspace{1em}$ [32,50] $ & $Q_{8}\circ D_{4}$ & $ [F_{1},F_{2},F_{4},F_{3} F_{1} F_{2} F_{3} F_{4} F_{3}^{-1} F_{2}^{-1} F_{1}^{-1} F_{3}^{-1}] $ \\ 
 &&\\ 
\textbf{45:} \hspace{1em}$ [32,35] $ & $\mathbb{Z}_{4}\rtimes Q_{8}$ & $ [F_{1}^{-2},F_{2} F_{1}^{-1},F_{4} F_{1}^{-1},F_{3} F_{2} F_{1}^{-1} F_{3}^{-1}] $ \\ 
 &&\\ 
\textbf{46:} \hspace{1em}$ [32,35] $ & $\mathbb{Z}_{4}\rtimes Q_{8}$ & $ [F_{1}^{-2},F_{4} F_{2}^{-1},F_{3} F_{2} F_{1}^{-1} F_{3}^{-1},F_{3} F_{4} F_{1}^{-1} F_{3}^{-1}] $ \\ 
 &&\\ 
\textbf{47:} \hspace{1em}$ [32,35] $ & $\mathbb{Z}_{4}\rtimes Q_{8}$ & $ [F_{1}^{-2},F_{4} F_{1}^{-1},F_{3} F_{2}^{-2} F_{3}^{-1},F_{3} F_{4} F_{1}^{-1} F_{3}^{-1},F_{1} F_{3} F_{2} F_{1}^{-1} F_{3}^{-1} F_{1}^{-1}] $ \\ 
 &&\\ 
\textbf{48:} \hspace{1em}$ [32,8] $ & $\mathbb{Z}_{4}\cdot D_{4}$ & $ [F_{1}^{-2},F_{2} F_{1}^{-1},F_{4} F_{1}^{-1},F_{1} F_{3} F_{2} F_{4}^{-1} F_{3}^{-1}] $ \\ 
 &&\\ 
\rowcolor{LightSteelBlue}\textbf{49:} \hspace{1em}$ [32,11] $ & $\mathbb{Z}_{4}\wr \mathbb{Z}_{2}$ & $ [F_{1}^{-2},F_{2} F_{3} F_{2}^{-1},F_{3} F_{4} F_{3}^{-1},F_{1} F_{2} F_{4} F_{3}^{-1} F_{1}^{-1}] $ \\ 
 &&\\ 
\rowcolor{LightSteelBlue}\textbf{50:} \hspace{1em}$ [32,11] $ & $\mathbb{Z}_{4}\wr \mathbb{Z}_{2}$ & $ [F_{1}^{-2},F_{1} F_{4} F_{3}^{-1},F_{1} F_{2} F_{3} F_{2}^{-1} F_{1}^{-1}] $ \\ 
 &&\\ 
\textbf{51:} \hspace{1em}$ [32,44] $ & $\mathbb{D}_{4}\rtimes \mathbb{Z}_{2}$ & $ [F_{1},F_{2},F_{3} F_{2} F_{1}^{-1} F_{3}^{-1},F_{4} F_{3} F_{2} F_{1}^{-1} F_{3}^{-1} F_{4}^{-1}] $ \\ 
 &&\\ 
\rowcolor{LightSteelBlue}\textbf{52:} \hspace{1em}$ [32,11] $ & $\mathbb{Z}_{4}\wr \mathbb{Z}_{2}$ & $ [F_{1},F_{2},F_{3} F_{1} F_{2} F_{3}^{-1} F_{4}^{-1}] $ \\ 
 &&\\ 
\textbf{53:} \hspace{1em}$ [32,44] $ & $\mathbb{D}_{4}\rtimes \mathbb{Z}_{2}$ & $ [F_{1},F_{4} F_{2}^{-1},F_{3} F_{4} F_{2}^{-1} F_{3}^{-1}] $ \\ 
 &&\\ 
\textbf{54:} \hspace{1em}$ [32,44] $ & $\mathbb{D}_{4}\rtimes \mathbb{Z}_{2}$ & $ [F_{1},F_{4},F_{3} F_{4} F_{1}^{-1} F_{3}^{-1},F_{2} F_{3} F_{4} F_{1}^{-1} F_{3}^{-1} F_{2}^{-1}] $ \\ 
 &&\\ 
\textbf{55:} \hspace{1em}$ [48,32] $ & $\mathbb{Z}_{2}\times \mathbb{T}$ & $ [F_{1}^{-2},F_{2} F_{1}^{-1},F_{3} F_{1}^{-1},F_{4} F_{3} F_{2} F_{1}^{-1} F_{3}^{-1} F_{4}^{-1}] $ \\ 
 &&\\ 
\textbf{56:} \hspace{1em}$ [48,32] $ & $\mathbb{Z}_{2}\times \mathbb{T}$ & $ [F_{1}^{-2},F_{4} F_{1}^{-1},F_{2} F_{3} F_{2}^{-1} F_{1}^{-1},F_{3} F_{4} F_{1}^{-1} F_{3}^{-1}] $ \\ 
 &&\\ 
\textbf{57:} \hspace{1em}$ [48,32] $ & $\mathbb{Z}_{2}\times \mathbb{T}$ & $ [F_{1}^{-2},F_{4} F_{2}^{-1},F_{2} F_{3} F_{2}^{-1} F_{1}^{-1}] $ \\ 
 &&\\ 
\textbf{58:} \hspace{1em}$ [48,29] $ & $\mathbb{T}\rtimes \mathbb{Z}_{2}$ & $ [F_{1}^{-2},F_{2}^{-2},F_{1} F_{2} F_{4}^{-1},F_{1} F_{3} F_{2}^{-1}] $ \\ 
 &&\\ 
\textbf{59:} \hspace{1em}$ [48,28] $ & $\mathbb{O}$ & $ [F_{1}^{-2},F_{4},F_{1} F_{3} F_{1}^{-1},F_{2} F_{3} F_{2}^{-1}] $ \\ 
 &&\\ 
\textbf{60:} \hspace{1em}$ [48,29] $ & $\mathbb{T}\rtimes \mathbb{Z}_{2}$ & $ [F_{1}^{-2},F_{2}^{-2},F_{3} F_{2}^{-1},F_{1} F_{2} F_{4}^{-1}] $ \\ 
 &&\\ 
\textbf{61:} \hspace{1em}$ [48,28] $ & $\mathbb{O}$ & $ [F_{1}^{-2},F_{2},F_{3},F_{1} F_{3} F_{4} F_{3}^{-1} F_{1}^{-1}] $ \\ 
 &&\\ 
\textbf{62:} \hspace{1em}$ [48,29] $ & $\mathbb{T}\rtimes \mathbb{Z}_{2}$ & $ [F_{1}^{-2},F_{2}^{-2},F_{1} F_{2} F_{4}^{-1},F_{2} F_{3} F_{4}^{-1}] $ \\ 
 &&\\ 
\textbf{63:} \hspace{1em}$ [48,28] $ & $\mathbb{O}$ & $ [F_{1},F_{2} F_{3} F_{2}^{-1},F_{3} F_{4} F_{3}^{-1}] $ \\ 
 &&\\ 
\textbf{64:} \hspace{1em}$ [48,32] $ & $\mathbb{Z}_{2}\times \mathbb{T}$  & $ [F_{1}^{-2},F_{1} F_{4} F_{3}^{-1} F_{1}^{-1},F_{2} F_{3} F_{2}^{-1} F_{1}^{-1}] $ \\ 
 &&\\ 
\textbf{65:} \hspace{1em}$ [64,137] $ & $D_{4}\cdot D_{4}$ & $ [F_{1},F_{2},F_{4},F_{3} F_{2} F_{1}^{-1} F_{3}^{-1}] $ \\ 
 &&\\ 
\textbf{66:} \hspace{1em}$ [64,37] $ & $\mathbb{Z}_{4}^{2}\cdot \mathbb{Z}_{4}$ & $ [F_{1}^{-2},F_{2} F_{1}^{-1},F_{4} F_{1}^{-1},F_{3} F_{2} F_{1}^{-1} F_{3}^{-1},F_{1} F_{3} F_{1} F_{4}^{-1} F_{3}^{-1}] $ \\ 
 &&\\ 
\textbf{67:} \hspace{1em}$ [64,137] $ & $D_{4}\cdot D_{4}$ & $ [F_{1}^{-2},F_{4} F_{2}^{-1},F_{1} F_{3} F_{1}^{-1},F_{2} F_{3} F_{2}^{-1}] $ \\ 
 &&\\ 
\textbf{68:} \hspace{1em}$ [64,37] $ & $\mathbb{Z}_{4}^{2}\cdot \mathbb{Z}_{4}$ & $ [F_{1}^{-2},F_{4} F_{2}^{-1},F_{1} F_{3} F_{2}^{-1}] $ \\ 
 &&\\ 
\textbf{69:} \hspace{1em}$ [64,137] $ & $D_{4}\cdot D_{4}$ & $ [F_{1}^{-2},F_{3},F_{4} F_{1}^{-1},F_{1} F_{2} F_{3} F_{2}^{-1} F_{1}^{-1}] $ \\ 
 &&\\ 
\textbf{70:} \hspace{1em}$ [64,37] $ & $\mathbb{Z}_{4}^{2}\cdot \mathbb{Z}_{4}$ & $ [F_{1}^{-2},F_{2}^{-2},F_{4} F_{1}^{-1},F_{1} F_{2} F_{3}^{-1}] $ \\ 
 &&\\ 
\textbf{71:} \hspace{1em}\textcolor{DarkSalmon}{$\bigstar$}$ [96,190] $\textcolor{DarkSalmon}{$\bigstar$} & $Q_{8}\cdot D_{6}$ & $ [F_{1},F_{2},F_{3},F_{4} F_{3} F_{2} F_{1}^{-1} F_{3}^{-1} F_{4}^{-1}] $ \\ 
 &&\\ 
\textbf{72:} \hspace{1em}\textcolor{DarkSalmon}{$\bigstar$}$ [96,190] $\textcolor{DarkSalmon}{$\bigstar$} & $Q_{8}\cdot D_{6}$ & $ [F_{1},F_{4},F_{2} F_{3} F_{2}^{-1},F_{3} F_{4} F_{1}^{-1} F_{3}^{-1}] $ \\ 
 &&\\ 
\textbf{73:} \hspace{1em}\textcolor{DarkSalmon}{$\bigstar$}$ [96,190] $\textcolor{DarkSalmon}{$\bigstar$} & $Q_{8}\cdot D_{6}$ & $ [F_{1},F_{4} F_{2}^{-1},F_{2} F_{3} F_{2}^{-1}] $ \\ 
 &&\\ 
\rowcolor{LightSteelBlue}\textbf{74:} \hspace{1em}$ [128,937] $ & $Q_{8}\wr \mathbb{Z}_{2}$ & $ [F_{1},F_{2},F_{4},F_{3} F_{2} F_{1}^{-1} F_{3}^{-1},F_{3} F_{4} F_{1}^{-1} F_{3}^{-1}] $ \\ 
 &&\\ 
\textbf{75:} \hspace{1em}$ [192,1022] $ & $Q_{8}\rtimes \mathbb{T}$ & $ [F_{1}^{-2},F_{2} F_{1}^{-1},F_{3} F_{1}^{-1},F_{4} F_{1}^{-1}] $ \\ 
 &&\\ 
\textbf{76:} \hspace{1em}$ [384,18130] $ & $(Q_{8}^{2}\rtimes \mathbb{Z}_{3})\rtimes \mathbb{Z}_{2}$ & $ [F_{1},F_{2},F_{3},F_{4}]$\\

& &\\
\hline

\caption{All subgroups of $G_{1}$ up to conjugacy (except those described at the beginning of the subsection). Groups acting in a representation such that the linear quotient is known to admit a symplectic resolution are highlighted. Those for which it is not yet known are marked by stars. The group $[96,190]$ is labelled as $(\mu_{4} \mid \mu_{2}; \textbf{O}\mid \textbf{T})$ in \S 3 of \cite{cohen1980finite}.}
\end{longtable}

\end{scriptsize}

\vspace{5em}

\begin{table}[p]
\begin{scriptsize}
\begin{center}
\scriptsize{
    
    \begin{tabular}{|l|l|l|}
    \hline
    \multicolumn{3}{|c|}{}\\
    \multicolumn{3}{|c|}{\large{\textbf{Subgroups of $G_{2}$}}}\\
    \multicolumn{3}{|c|}{}\\
    \hline
    & &\\ 
    Group ID & Name & Generators \\
    & &\\
    \hline
    & &\\
    \textbf{1:} \hspace{1em}$ [8,4] $ & $Q_{8}$ & $ [F_{3}^{-1} F_{2}^{-1},F_{1} F_{3} F_{1} F_{4}^{-1} F_{2} F_{1}^{-1}] $ \\ 
 &&\\ 
\textbf{2:} \hspace{1em}$ [8,4] $ & $Q_{8}$ & $ [F_{1}^{3},F_{2} F_{4} F_{3}^{-1} F_{1},F_{3} F_{1} F_{4}^{-1} F_{2}] $ \\ 
 &&\\ 
\textbf{3:} \hspace{1em}$ [12,1] $ & $\mathbb{D}_{3}$ & $ [F_{1}^{3},F_{1} F_{2} F_{3}^{-1},F_{2} F_{4} F_{3}^{-1} F_{1}^{-1}] $ \\ 
 &&\\ 
\textbf{4:} \hspace{1em}$ [12,1] $ & $\mathbb{D}_{3}$ & $ [F_{1}^{3},F_{1} F_{2} F_{3}^{-1},F_{2} F_{4} F_{3} F_{2}^{-1}] $ \\ 
 &&\\ 
\textbf{5:} \hspace{1em}$ [16,9] $ & $\mathbb{D}_{4}$ & $ [F_{3}^{-1} F_{2}^{-1},F_{1}^{-1} F_{2} F_{3} F_{1}^{-1},F_{2} F_{4} F_{3}^{-1} F_{1}] $ \\ 
 &&\\ 
\textbf{6:} \hspace{1em}$ [20,1] $ & $\mathbb{D}_{5}$ & $ [F_{2}^{-1} F_{1}^{-1},F_{3} F_{1}] $ \\ 
 &&\\ 
\textbf{7:} \hspace{1em}$ [24,3] $ & $\mathbb{T}$ & $ [F_{1},F_{2} F_{4} F_{3}^{-1}] $ \\ 
 &&\\ 
\rowcolor{LightSteelBlue}\textbf{8:} \hspace{1em}$ [24,3] $ & $\mathbb{T}$ & $ [F_{3}^{-1} F_{2}^{-1},F_{1}^{3},F_{1}^{-1} F_{2} F_{3} F_{1}^{-1},F_{2} F_{4} F_{1} F_{2}^{-1}] $ \\ 
 &&\\ 
\textbf{9:} \hspace{1em}$ [36,7] $ & $\mathbb{Z}_{3}\rtimes \mathbb{D}_{3}$ & $ [F_{2}^{-1} F_{1}^{-1},F_{3},F_{1}^{-1} F_{3}^{-1} F_{2}^{-1} F_{1},F_{2} F_{4} F_{3}^{-1} F_{1}^{-1}] $ \\ 
 &&\\ 
\textbf{10:} \hspace{1em}$ [48,28] $ & $\mathbb{O}$ & $ [F_{1},F_{3}^{-1} F_{2}^{-1},F_{2} F_{4} F_{3}^{-1}] $ \\ 
 &&\\ 
\textbf{11:} \hspace{1em}$ [48,28] $ & $\mathbb{O}$ & $ [F_{3} F_{2},F_{3}^{-1} F_{2}^{-1},F_{1}^{3},F_{1} F_{4} F_{3} F_{1}^{-1},F_{1} F_{4}^{-1} F_{3}^{-1} F_{1}^{-1},F_{1}^{-1} F_{2} F_{3} F_{1}^{-1}] $ \\ 
 &&\\ 
\textbf{12:} \hspace{1em}$ [72,19] $ & $\mathbb{Z}_{3}^{2}\rtimes \mathbb{Z}_{8}$ & $ [F_{2}^{-1} F_{1}^{-1},F_{3},F_{1}^{-1} F_{2} F_{4}^{-1} F_{1}^{-1}] $ \\ 
 &&\\ 
\textbf{13:} \hspace{1em}$ [120,5] $ & $\mathbb{I}$ & $ [F_{2}^{-1} F_{4}^{-1} F_{1}^{3} F_{3} F_{2} F_{4} F_{3} F_{2} F_{1}^{3} F_{4} F_{2},F_{1}^{5} F_{3} F_{2} F_{4} F_{3} F_{2} F_{1}] $ \\ 
 &&\\ 
\textbf{14:} \hspace{1em}\textcolor{DarkSalmon}{$\bigstar$}$ [120,5] $\textcolor{DarkSalmon}{$\bigstar$} & $\mathbb{I}$ & $ [F_{2}^{-1} F_{4}^{-2} F_{1}^{4} F_{2} F_{3} F_{2} F_{1} F_{4} F_{3} F_{2} F_{1}^{3} F_{4}^{2} F_{2},F_{1}^{5} F_{2} F_{3} F_{2} F_{1} F_{4} F_{3} F_{2} F_{1}^{2}] $ \\ 
 &&\\ 
\textbf{15:} \hspace{1em}\textcolor{DarkSalmon}{$\bigstar$}$ [720,409] $\textcolor{DarkSalmon}{$\bigstar$} & --- & $ [F_{1},F_{2},F_{3},F_{4}] $\\
    & &\\
    \hline
    \end{tabular}
}
\end{center}
\end{scriptsize}
\caption{All subgroups of $G_{2}$ up to conjugacy (except those described at the beginning of the subsection). Groups acting in a representation such that the linear quotient is known to admit a symplectic resolution are highlighted. Those for which it is not yet known are marked by stars. The group $[120,5]$ is labelled as $W(O_{1})$, and the group $[720,409]$ as $W(O_{2})$ in \S 4 of \cite{cohen1980finite}.}
\label{tab:table 2}
\end{table}

\newpage
\begin{scriptsize}
\begin{longtable}{|l|l|l|}
    \hline
    \multicolumn{3}{|c|}{}\\
    \multicolumn{3}{|c|}{\large{\textbf{Subgroups of $G_{3}$}}}\\
    \multicolumn{3}{|c|}{}\\
    \hline
    & &\\
     Group ID & Name & Generators \label{tab:table 3}\\
    & &\\
    \hline 
    & &\\
    \textbf{1:} $ \qquad[8,4] $ & $Q_{8}$ & $ [F_{4} F_{3}^{-1},F_{1}^{-1} F_{4}^{-1} F_{3} F_{1}] $ \\ 
 &&\\ 
\rowcolor{LightSteelBlue}\textbf{2:} $ \qquad [8,3] $ & $D_{4}$ & $ [F_{4}^{-1} F_{3},F_{1}^{-1} F_{3} F_{4} F_{3} F_{1}^{-1} F_{4}^{-1} F_{3} F_{1}^{-1}] $ \\ 
 &&\\ 
\textbf{3:} $ \qquad [8,4] $ & $Q_{8}$ & $ [F_{1} F_{4} F_{1} F_{3} F_{1},F_{1} F_{4}^{-1} F_{1} F_{3}^{-1} F_{1}^{-2}] $ \\ 
 &&\\ 
\textbf{4:} $ \qquad [8,4] $ & $Q_{8}$ & $ [F_{1}^{3},F_{1} F_{4} F_{1}^{-1} F_{4}^{-1} F_{1},(F_{3} F_{1})^{2} F_{4} F_{1}^{-1} F_{3}] $ \\ 
 &&\\ 
\textbf{5:} $ \qquad [8,4] $ & $Q_{8}$ & $ [F_{1}^{3},F_{3} F_{1} F_{3}^{-1} F_{1},F_{3}^{-1} F_{1}^{-1} F_{4}^{-1} F_{3} F_{1} F_{3}] $ \\ 
 &&\\ 
\textbf{6:} $ \qquad [12,1] $ & $\mathbb{D}_{3}$ & $ [F_{1}^{3},F_{3} F_{1} F_{3}^{-1},F_{1} F_{3} F_{1}^{-1} F_{4}^{-1} F_{3}^{-1}] $ \\ 
 &&\\ 
\textbf{7:} $ \qquad [16,12] $ & $\mathbb{Z}_{2}\times Q_{8}$ & $ [F_{4} F_{3}^{-1},F_{1} F_{4}^{-1} F_{3} F_{1}^{-1},F_{1}^{-1} F_{4}^{-1} F_{3} F_{1}] $ \\ 
 &&\\ 
\textbf{8:} $ \qquad [16,12] $ & $\mathbb{Z}_{2}\times Q_{8}$ & $ [F_{1}^{3},(F_{1} F_{4})^{2} F_{1},(F_{1} F_{4}^{-1})^{2} F_{1},F_{1}^{-1} F_{3} F_{1}^{-1} F_{4}^{-1} F_{3}^{-1} F_{1}^{-1}] $ \\ 
 &&\\ 
\textbf{9:} $ \qquad [16,13] $ & $\mathbb{Z}_{4}\circ D_{4}$ & $ [F_{4}^{-1} F_{3},F_{1}^{-1} F_{4}^{-1} F_{3} F_{1},F_{1} F_{3} F_{4} F_{3} F_{1}^{-1}] $ \\ 
 &&\\ 
\textbf{10:} $ \qquad [16,12] $ & $\mathbb{Z}_{2}\times Q_{8}$ & $ [F_{1}^{3},F_{1}^{-1} F_{4}^{-1} F_{3} F_{1},F_{1} F_{3} F_{4} F_{3} F_{1}^{-1},(F_{1} F_{3}^{-1})^{2} F_{1}] $ \\ 
 &&\\ 
\textbf{11:} $ \qquad [16,6] $ & $M_{4}(2)$ & $ [F_{1}^{3},F_{1}^{-1} F_{4}^{-1} F_{3} F_{1},F_{1} F_{3} F_{4} F_{3} F_{1}^{-1},F_{3} F_{1}^{-1} F_{4} F_{1}^{-1} F_{3}] $ \\ 
 &&\\ 
\textbf{12:} $ \qquad [16,4] $ & $\mathbb{Z}_{4}\rtimes \mathbb{Z}_{4}$ & $ [F_{1}^{3},(F_{1} F_{4})^{2} F_{1},F_{1} F_{4}^{-1} F_{1} F_{3}^{-1} F_{1}] $ \\ 
 &&\\ 
\textbf{13:} $ \qquad [16,4] $ & $\mathbb{Z}_{4}\rtimes \mathbb{Z}_{4}$ & $ [F_{4} F_{3}^{-1},F_{1} F_{4}^{-1} F_{1} F_{3}^{-1} F_{1}] $ \\ 
 &&\\ 
\textbf{14:} $ \qquad [16,4] $ & $\mathbb{Z}_{4}\rtimes \mathbb{Z}_{4}$ & $ [F_{1}^{3},F_{1} F_{4}^{-1} F_{3} F_{1}^{-1},(F_{1} F_{3}^{-1})^{2} F_{1}] $ \\ 
 &&\\ 
\textbf{15:} $ \qquad [16,8] $ & $QD_{16}$ & $ [F_{4}^{-1} F_{3},F_{3} F_{1}^{-1} F_{3} F_{1},F_{1} F_{4} F_{1}^{-1} F_{3} F_{1}^{-1} F_{4}^{-1} F_{3}^{-1} F_{1}^{-1}] $ \\ 
 &&\\ 
\textbf{16:} $ \qquad [16,9] $ & $\mathbb{D}_{4}$ & $ [F_{1}^{3},F_{3} F_{1} F_{3}^{-1} F_{1},F_{3} F_{4} F_{3}] $ \\ 
 &&\\ 
\textbf{17:} $ \qquad [20,1] $ & $\mathbb{D}_{5}$ & $ [F_{1}^{3},F_{3} F_{1} F_{3} F_{1}^{-1},F_{3}^{-1} F_{1} F_{4}^{-1}] $ \\ 
 &&\\ 
\textbf{18:} $ \qquad [24,3] $ & $\mathbb{T}$ & $ [F_{4} F_{3}^{-1},F_{1}^{-1} F_{3} F_{1}] $ \\ 
 &&\\ 
\textbf{19:} $ \qquad [24,3] $ & $\mathbb{T}$ & $ [F_{1}^{-1} F_{4} F_{1}^{-2},F_{3} F_{1}^{-2} F_{3}^{-1}] $ \\ 
 &&\\ 
\textbf{20:} $ \qquad [24,1] $ & $\mathbb{Z}_{3}\rtimes \mathbb{Z}_{8}$ & $ [F_{1}^{3},F_{3} F_{1} F_{3}^{-1},F_{3} F_{4} F_{3},F_{1} F_{3}^{-1} F_{4} F_{1} F_{3} F_{1} F_{4}] $ \\ 
 &&\\ 
\textbf{21:} $ \qquad [24,4] $ & $\mathbb{D}_{6}$ & $ [F_{1}^{3},F_{3} F_{1} F_{3}^{-1},F_{3} F_{4} F_{3},F_{3}^{-1} F_{1} F_{3}^{-1} F_{1}^{-1}] $ \\ 
 &&\\ 
\textbf{22:} $ \qquad [24,10] $ & $\mathbb{Z}_{3}\times D_{4}$ & $ [F_{1}^{3},F_{3} F_{1} F_{3}^{-1},F_{3} F_{4} F_{3},F_{1}^{-1} F_{4} F_{1}^{-1} F_{3} F_{4} F_{1}^{-1}] $ \\ 
 &&\\ 
\textbf{23:} $ \qquad [24,3] $ & $\mathbb{T}$ & $ [F_{1}^{3},F_{3} F_{1} F_{3}^{-1},F_{1} F_{3} F_{1}^{-1} F_{3} F_{4}] $ \\ 
 &&\\ 
\rowcolor{LightSteelBlue}\textbf{24:} $ \qquad [32,50] $ & $Q_{8}\circ D_{4}$ & $ [F_{4} F_{3}^{-1},F_{4}^{-1} F_{3},F_{1} F_{4}^{-1} F_{3} F_{1}^{-1},F_{1}^{-1} F_{4}^{-1} F_{3} F_{1}] $ \\ 
 &&\\ 
\textbf{25:} $ \qquad [32,35] $ & $\mathbb{Z}_{4}\rtimes Q_{8}$ & $ [F_{4} F_{3}^{-1},F_{1} F_{4}^{-1} F_{3} F_{1}^{-1},F_{1}^{-1} F_{4}^{-1} F_{3} F_{1},(F_{1} F_{3}^{-1})^{2} F_{1}] $ \\ 
 &&\\ 
\textbf{26:} $ \qquad [32,8] $ & $\mathbb{Z}_{4}\cdot D_{4}$ & $ [F_{4} F_{3}^{-1},F_{1} F_{4}^{-1} F_{3} F_{1}^{-1},F_{1}^{-1} F_{4}^{-1} F_{3} F_{1},F_{3} F_{1} F_{4}^{-1} F_{3} F_{1}^{-1} F_{3}^{-1},(F_{3} F_{1}^{-1})^{2} F_{4}] $ \\ 
 &&\\ 
\rowcolor{LightSteelBlue}\textbf{27:} $ \qquad [32,11] $ & $\mathbb{Z}_{4}\wr \mathbb{Z}_{2}$ & $ [F_{4}^{-1} F_{3},F_{1}^{-1} F_{4}^{-1} F_{3} F_{1},F_{1} F_{3} F_{4} F_{3} F_{1}^{-1},F_{1} F_{3}^{-1} F_{1} F_{4}^{-1} F_{1}] $ \\ 
 &&\\ 
\textbf{28:} $ \qquad [32,35] $ & $\mathbb{Z}_{4}\rtimes Q_{8}$ & $ [F_{4} F_{3}^{-1},F_{1} F_{3} F_{1} F_{4} F_{1},F_{1} F_{4}^{-1} F_{1} F_{3}^{-1} F_{1}] $ \\ 
 &&\\ 
\textbf{29:} $ \qquad [32,44] $ & $\mathbb{D}_{4}\rtimes \mathbb{Z}_{2}$ & $ [F_{1}^{3},F_{1}^{-1} F_{4}^{-1} F_{3} F_{1},F_{3} F_{4} F_{3},F_{1} F_{3} F_{4} F_{3} F_{1}^{-1},(F_{1} F_{3}^{-1})^{2} F_{1}] $ \\ 
 &&\\ 
\textbf{30:} $ \qquad [48,33] $ & $\mathbb{Z}_{4}\circ \mathbb{T}$ & $ [F_{4}^{-1} F_{3},F_{1}^{-1} F_{4}^{-1} F_{3} F_{1},F_{3} F_{1}^{-1} F_{3}^{-1} F_{1},F_{1} F_{3} F_{4} F_{3} F_{1}^{-1}] $ \\ 
 &&\\ 
\textbf{31:} $ \qquad [48,32] $ & $\mathbb{Z}_{2}\times \mathbb{T}$ & $ [F_{4} F_{3}^{-1},F_{1} F_{4}^{-1} F_{3} F_{1}^{-1},F_{1}^{-1} F_{4}^{-1} F_{3} F_{1},F_{3} F_{1} F_{3}^{-1}] $ \\ 
 &&\\ 
\textbf{32:} $ \qquad [48,28] $ & $\mathbb{O}$ & $ [F_{4} F_{3}^{-1},F_{1}^{-1} F_{3} F_{1},F_{3}^{-1} F_{1} F_{4}^{-1} F_{3}^{-1} F_{1}^{-1} F_{3}^{-1}] $ \\ 
 &&\\ 
\textbf{33:} $ \qquad [48,32] $ & $\mathbb{Z}_{2}\times \mathbb{T}$ & $ [F_{1}^{3},F_{1} F_{4}^{-1} F_{3}^{-1} F_{1}^{-1},F_{1}^{-1} F_{4} F_{1}] $ \\ 
 &&\\ 
\textbf{34:} $ \qquad [48,16] $ & $\mathbb{D}_{6}\rtimes \mathbb{Z}_{2}$ & $ [F_{1}^{3},F_{3} F_{1} F_{3}^{-1},F_{3} F_{4} F_{3},F_{3}^{-1} F_{1} F_{3}^{-1} F_{1}^{-1},F_{1}^{-1} F_{4} F_{1}^{-1} F_{3} F_{4} F_{1}^{-1}] $ \\ 
 &&\\ 
\textbf{35:} $ \qquad [48,32] $ & $\mathbb{Z}_{2}\times \mathbb{T}$ & $ [F_{1}^{3},F_{1} F_{4} F_{3} F_{1}^{-1},F_{1}^{-1} F_{4} F_{3} F_{1},F_{1} F_{3}^{-1} F_{1} F_{4} F_{1}] $ \\ 
 &&\\ 
\textbf{36:} $ \qquad [64,137] $ & $D_{4}\cdot D_{4}$ & $ [F_{4} F_{3}^{-1},F_{4}^{-1} F_{3},F_{1} F_{4}^{-1} F_{3} F_{1}^{-1},F_{1}^{-1} F_{4}^{-1} F_{3} F_{1},(F_{1} F_{3}^{-1})^{2} F_{1}] $ \\ 
 &&\\ 
\textbf{37:} $ \qquad [64,37] $ & $\mathbb{Z}_{4}^{2}\cdot \mathbb{Z}_{4}$ & $ [F_{4} F_{3}^{-1},F_{1} F_{4}^{-1} F_{3} F_{1}^{-1},F_{1}^{-1} F_{4}^{-1} F_{3} F_{1},(F_{1} F_{3}^{-1})^{2} F_{1},(F_{3} F_{1})^{2} F_{4}] $ \\ 
 &&\\ 
\textbf{38:} $ \qquad [96,202] $ & $D_{4}\circ \mathbb{T}$ & $ [F_{4} F_{3}^{-1},F_{4}^{-1} F_{3},F_{1} F_{4}^{-1} F_{3} F_{1}^{-1},F_{1}^{-1} F_{4}^{-1} F_{3} F_{1},F_{3} F_{1} F_{3}^{-1}] $ \\ 
 &&\\ 
\textbf{39:} $ \qquad [96,191] $ & $\mathbb{O}\rtimes \mathbb{Z}_{2}$ & $ [F_{4}^{-1} F_{3},F_{1}^{-1} F_{4}^{-1} F_{3} F_{1},F_{3} F_{1} F_{3} F_{1}^{-1},F_{3} F_{1}^{-1} F_{3}^{-1} F_{1},F_{1} F_{3} F_{4} F_{3} F_{1}^{-1}] $ \\ 
 &&\\ 
\textbf{40:} $ \qquad [96,67] $ & $\mathbb{T}\rtimes \mathbb{Z}_{4}$ & $ [F_{4}^{-1} F_{3},F_{1}^{-1} F_{4}^{-1} F_{3} F_{1},F_{3} F_{1} F_{4} F_{1}^{-1},F_{3} F_{1}^{-1} F_{3}^{-1} F_{1}] $ \\ 
 &&\\ 
\textbf{41:} $ \qquad [120,5] $ & $\mathbb{I}$ & $ [F_{3}^{-2} F_{1}^{-1} F_{4}^{2} F_{1} F_{3} F_{1}^{-1} F_{4} F_{1} F_{3}^{2},F_{4}^{2} F_{1} F_{3} F_{1}^{-1} F_{4}] $ \\ 
 &&\\ 
\rowcolor{LightSteelBlue}\textbf{42:} $ \qquad [128,937] $ & $Q_{8}\wr \mathbb{Z}_{2}$ & $ [F_{4} F_{3}^{-1},F_{4}^{-1} F_{3},F_{1} F_{4}^{-1} F_{3} F_{1}^{-1},F_{1}^{-1} F_{4}^{-1} F_{3} F_{1},(F_{1} F_{3})^{2} F_{1},(F_{1} F_{3}^{-1})^{2} F_{1}] $ \\ 
 &&\\ 
\textbf{43:} $ \qquad [160,199] $ & $(Q_{8}\circ D_{4})\rtimes \mathbb{Z}_{5}$ & $ [F_{4} F_{3}^{-1},F_{4}^{-1} F_{3},F_{1} F_{3} F_{1}] $ \\ 
 &&\\ 
\textbf{44:} $ \qquad [192,1022] $ & $Q_{8}\rtimes \mathbb{T}$ & $ [F_{4} F_{3}^{-1},F_{1} F_{3} F_{1}^{-1},F_{1}^{-1} F_{3} F_{1}] $ \\ 
 &&\\ 
\textbf{45:} \hspace{1em}\textcolor{DarkSalmon}{$\bigstar$}$  [192,989] $\textcolor{DarkSalmon}{$\bigstar$} & $D_{4}\cdot S_{4}$ & $ [F_{4} F_{3}^{-1},F_{4}^{-1} F_{3},F_{1} F_{4}^{-1} F_{3} F_{1}^{-1},F_{1}^{-1} F_{4}^{-1} F_{3} F_{1},F_{3} F_{1} F_{3}^{-1},F_{3}^{-1} F_{1} F_{3}^{-1} F_{1}^{-1}] $ \\ 
 &&\\ 
\textbf{46:}\hspace{1em} \textcolor{DarkSalmon}{$\bigstar$}$ [320,1581] $\textcolor{DarkSalmon}{$\bigstar$} & $(Q_{8}\circ D_{4})\cdot D_{5}$ & $ [F_{4} F_{3}^{-1},F_{4}^{-1} F_{3},F_{1} F_{3} F_{1},F_{3} F_{1} F_{3} F_{1}^{-1}] $ \\ 
 &&\\ 
\textbf{47:}$\qquad  [384,618] $ & $(Q_{8}^{2}\rtimes \mathbb{Z}_{2})\rtimes \mathbb{Z}_{3}$ & $ [F_{4} F_{3}^{-1},F_{4}^{-1} F_{3},F_{1} F_{3} F_{1}^{-1},F_{1}^{-1} F_{3} F_{1}] $ \\ 
 &&\\ 
\textbf{48:}\hspace{1em}\textcolor{DarkSalmon}{$\bigstar$}$ [1920,241003] $\textcolor{DarkSalmon}{$\bigstar$} & --- & $ [F_{1},F_{3},F_{4}] $ \\
     & &\\
     \hline
    \caption{ All subgroups of $G_{3}$ up to conjugacy (except those described at the beginning of the subsection). Groups acting in a representation such that the linear quotient is known to admit a symplectic resolution are highlighted. Those for which it is not yet known are marked by stars. The group $[192,989]$ is labelled as $(\mu_{8} \mid \mu_{4}; \textbf{O}\mid \textbf{T})$ in \S 3 of \cite{cohen1980finite}. The groups $[320,1581]$ and $[1920,241003]$ are labelled as $W(P_{1})$ and $W(P_{2})$ in \S 4 of \cite{cohen1980finite}.}
    \end{longtable}
\end{scriptsize}

\section{Studying the Singular Sets}\label{ref: studying singular sets}

The above classification lists all finite subgroups $H$ of $\op{Sp}(2)$ which act on the maximal torus $T^{8} = \mathbb{C}^{4}/\Lambda_{E_{8}}$ of $E_{8}$ preserving a hyperK\"ahler structure. Therefore, $T^{8}/H$ admits the structure of a hyperK\"ahler orbifold. However, this is not a complete classification of the finite hyperK\"ahler symmetries of $T^{8}$ since we should also consider adding translational isometries to the elements of $H$. Given a subgroup $H\subset \op{Sp}(2)$, one can consider adding translations to each generator of $H$ such that the group relations of $H$ are preserved when acting on $T^{8}$. Since any translation preserves the hyperK\"ahler structure, any such group also produces a hyperK\"ahler orbifold with holonomy group isomorphic to $H$. This is equivalent to choosing a group $\hat{H}\subset \op{O}\ltimes \mathbb{R}^{8}$ such that $\pi(\hat{H}) = H$ under the map $\pi:\op{O}(8)\ltimes \mathbb{R}^{8}\to \op{O}(8)$ and there is a subgroup $T_{\Lambda_{E_{8}}}\subset \hat{H}$ isomorphic to $\Lambda_{E_{8}}$. Then $\hat{H}$ is known as a space group with point group $H$ and,
\begin{equation}
    \frac{\mathbb{R}^{8}}{\hat{H}} \cong \frac{T^{8}}{H'},\quad \text{where }\; H' = \hat{H}/T_{E_{8}}\cong H
\end{equation}
In principle one can classify all possible space groups $\hat{H}$ with a given point group $H$ from one of the tables above. However, this turns out not to be necessary for deciding when $T^{8}/H'$ does not admit a symplectic resolution, for the majority of cases. Our main result, stated in theorem \ref{thm: theorem 1}, shows that none of the orbifolds $T^{8}/H'$ admit a symplectic resolution except for one case.

\begin{theorem}\label{thm: theorem 1}
    Let $T^{8} = \mathbb{C}^{4}/\Lambda_{E_{8}}$ be the maximal torus of $E_{8}$ and $G_{i}\subset \op{Aut}(\Lambda_{E_{8}})$ $(i=1,2,3)$ be as described in \ref{sec: mat reps}. Let  $H\subset G_{i}$ be a subgroup that is neither Abelian nor acting on $\mathbb{C}^{4}$ as $(\mathbb{C}^{2}/\Gamma_{ADE})\times (\mathbb{C}^{2}/\Gamma_{ADE}')$, then one can construct a group $\hat{H}\subset \op{O}(8)\ltimes \mathcal{T}(T^{8})$ (where $\mathcal{T}(T^{8})\cong \mathbb{R}^{8}$ is the normal subgroup of translations around the non-trivial cycles) such that under the map 
    \begin{equation}
    \pi: O(8)\ltimes A_{0}(T^{8})\to \op{O}(8),\quad (g,t)\mapsto g
    \end{equation}
    we have that $\pi(\hat{H}) = H$ and $\hat{H}$ contains a subgroup $T_{E_{8}} = \{(\mathbb{1}, v)\mid v\in \Lambda_{E_{8}}\}$. i.e. $\hat{H}$ is a choice of space group with point group $H$.
    
    For every $H$ (except for row 24 of table \ref{tab:table 3}) and for every choice of $\hat{H}$ the compact orbifold
    \begin{equation*}
        \frac{\mathbb{R}^{8}}{\hat{H}} \cong \frac{T^{8}}{H'},\quad H' = \hat{H}/T_{E_{8}}
    \end{equation*}
    always has at least one NRS singularity and thus is known not to admit a symplectic resolution to a compact hyperK\"ahler manifold. Row 24 of table \ref{tab:table 3} with the trivial choice of $\hat{H}$ (i.e. $\hat{H} = H\ltimes T_{E_{8}}$) gives rise to an orbifold identical to that discovered in \cite{donten201781} which admits a symplectic resolution diffeomorphic to $\op{Hilb}^{2}(K3)$.
\end{theorem}

In the following we outline the method used to arrive at the above result.

\subsection{Deciding non-resolvability}\label{sec: deciding non-resolvability}
In order to decide when a given orbifold $T^{8}/H'$ does not admit a symplectic resolution, we can apply the following method:

\begin{enumerate}
    \item Take one of the subgroups $H$ from the tables in section \ref{sec: list of subgroups} and choose a cyclic subgroup $\mathbb{Z}_{n} = \langle M\rangle$ of $H$, such that $M$ acts on $T^{8}$ with isolated fixpoints (i.e. such that $\det(\mathbb{1}-M)\neq 0$). Then for any choice of space group $\hat{H}$, it is always possible to conjugate $\hat{H}$ by a pure translation inside $\op{O}(8)\ltimes \mathcal{T}(T^{8})$ to get a new group $\hat{H}_{0}$, such that the chosen cyclic subgroup now acts with trivial translational part\footnote{Since, if $\det(\mathbb{1}-M)\neq 0$, the equation $(\mathbb{1},-r)(M,t)(\mathbb{1},r) = (M,t-(\mathbb{1}-M)r) \overset{!}{=}(M,\Vec{0})$ can be solved by $r = (\mathbb{1}-M)^{-1}t$.}.
    \item Then $\mathbb{R}^{8}/\hat{H}_{0}$ has an isolated $\mathbb{C}^{4}/K$ singularity, where $K\subset H$ is either $\mathbb{Z}_{n}$ or some other subgroup of $H$ containing it.

    \item If one can show that for all of the supergroups $K$ of the chosen cyclic subgroup $\mathbb{C}^{4}/K$ is an NRS singularity, then one can conclude that the singular set of $\mathbb{R}^{8}/\hat{H}_{0}$ will always have at least one NRS singularity. Since $\hat{H}_{0}$ and $\hat{H}$ are conjugate in $\op{O}(8)\ltimes \mathcal{T}(T^{8})$ the singular sets of $\mathbb{R}^{8}/\hat{H}$ and $\mathbb{R}^{8}/\hat{H}_{0}$ are identical, and so $\mathbb{R}^{8}/\hat{H}$ will also always have at least one NRS singularity. Note that this is true for any choice of space group $\hat{H}$ with point group $H$.
\end{enumerate}

To illustrate the method, we can look at the group with Small Group ID [32,8] which is an extension of $D_{4}$ by $\mathbb{Z}_{4}$, this occurs as a subgroup of $G_{1}$. This group is not on the list of resolvable singularities and it also has a $\mathbb{Z}_{8}$ subgroup acting with isolated fixpoints. The only supergroup chain that this $\mathbb{Z}_{8}$ fits into is
\begin{equation}
    \mathbb{Z}_{8}\subset M_{4}(2)\subset \op{SmallGroup(32,8)}
\end{equation}
Since these groups are not on the list of resolvable singularities we can rule out these examples because, depending on the translations, all such orbifolds will have either a $\mathbb{C}^{4}/\mathbb{Z}_{8}$ singularity, a $\mathbb{C}^{4}/M_{4}(2)$ or a $\mathbb{C}^{4}/\op{SmallGroup}(32,8)$ singularity. 

As a further example, take $\mathbb{Z}_{2}\times Q_{8}$ which occurs as a subgroup of $G_{1}$. Note that although this is of the form $\Gamma_{ADE}\times \Gamma_{ADE}'$, it is not acting as $(\mathbb{C}^{2}/\Gamma_{ADE})\times (\mathbb{C}^{2}/\Gamma_{ADE}')$ in this case and so it is possible to apply the above method (this can be seen by computing the characters). One can identify that there is a cyclic subgroup $\mathbb{Z}_{4}\subset \mathbb{Z}_{2}\times Q_{8}$ that is acting with isolated fixpoints. By studying the subgroup lattice of $\mathbb{Z}_{2}\times Q_{8}$, one can see that there are only the following possible supergroup chains,
\begin{align*}
        &\mathbb{Z}_{4}\subset Q_{8}\subset \mathbb{Z}_{2}\times Q_{8}\\
        \\
        &\mathbb{Z}_{4}\subset \mathbb{Z}_{2}\times \mathbb{Z}_{4}\subset \mathbb{Z}_{2}\times Q_{8}
\end{align*}
By checking characters, one can see that neither $Q_{8}$ or $\mathbb{Z}_{2}\times \mathbb{Z}_{4}$ are acting in the representation that admits a symplectic resolution, so all the groups in these chains are not on the list in section \ref{sec: symp resolutions} and this example can be ruled out too. Almost all of the groups in the tables of section \ref{sec: list of subgroups} can be ruled out in a similar vein. 

There are a small number of examples left that cannot be ruled out using this method. Namely, those examples that occur on the list of resolvable singularities in section \ref{sec: symp resolutions} and those examples that are not known to admit a symplectic resolution at present. The previous method then fails since every chain of supergroups containing the chosen cyclic subgroup will end with a group on the list of resolvable singularities or a group which may or may not be on the list. For these cases we must classify all possible translations and then investigate the singular sets. The remaining groups are those that are highlighted or starred in the tables in section \ref{sec: list of subgroups}.

In Appendix \ref{sec:class trans}, we classify all possible space groups compatible with these point groups and demonstrate that the singular sets have at least one NRS singularity in each case, except for one of the examples coming from $Q_{8}\circ D_{4}$ as already mentioned.

\vspace{1em}

\begin{remark}
Using our set of examples, the probability of finding a symplectically resolvable orbifold is less than 1\%. It would be interesting to know how reasonable a measure this is for the probability of finding resolvable examples in the set of all compact hyperK\"ahler 8-orbifolds.
\end{remark}

\section{Other Holonomy Groups}

One could also consider Kummer-type orbifolds with different holonomy groups and try to add translations to the group action such that all singularities can be desingularised to give a compact manifold with that holonomy group. For example, we can try to obtain compact Calabi-Yau manifolds from Calabi-Yau orbifolds. All the singularities of such an orbifold are locally of the form $\mathbb{C}^{n}/G$ for some $G\subset \op{SL}(n,\mathbb{C})$ (since every such $G$ is conjugate to a finite subgroup of $\op{SU}(n)$ one could also consider finite $G\subset \op{SU}(n)$). If all these singularities admit crepant resolutions \footnote{It can happen that if $X$ is a K\"ahler orbifold with non-isolated singularities, then some of the isolated singularities are not K\"ahler. See e.g. \S 6.6 of \cite{JoyceBook} for discussion of this. However, it is true that if $X$ admits crepant resolutions, at least one of them must be K\"ahler.}, we can desingularise the space and obtain a compact Calabi-Yau manifold. Analagously to the hyperK\"ahler case, we can make a list of all the singularities which are known to admit a crepant resolution and compare the singular set of our orbifold to this list. Here is a summary of what is known in each complex dimension $n$:
\begin{itemize}
    \item[$(n=2)$] Each singularity $\mathbb{C}^{2}/G$, for finite $G\subset \op{SU}(2)$, admits a \emph{unique} crepant resolution. These are know variously as Kleinian singularities, Du Val singularities or rational double points and have a close link to the Lie algebras of $ADE$-type through the McKay correspondence.
    \item[$(n=3)$] Each singularity $\mathbb{C}^{3}/G$, for finite $G\subset \op{SL}(3,\mathbb{C})$, admits a crepant resolution. However, it is not necessarily unique; there can be finitely many different crepant resolutions.
    \item[$(n\geq 4)$] Each singularity $\mathbb{C}^{4}/G$, for finite $G \subset \op{SL}(4,\mathbb{C})$ may or may not admit a crepant resolution. For example the singularity $\mathbb{C}^{4}/\{\pm \mathbb{1}\}$ is known to admit no crepant resolution. Many examples of finite $G\subset \op{SL}(4,\mathbb{C})$ that do have a crepant resolution are provided in \cite{hayashi2017existence}. For $n>4$, very little is known.
\end{itemize}

To illustrate this, take the following group $G = S_{3}\times \mathbb{Z}_{2}^{2}\subset \op{SL}(4,\mathbb{C})$ generated by the matrices,
\begin{equation}
    g_{1} = \begin{pmatrix}
        0 & 1 & & \\ 1&0&&\\&&-1&\\&&&1
    \end{pmatrix},\quad g_{2} = \begin{pmatrix}
        1 & -1 & & \\ 1&0&&\\&&1&\\&&&1
    \end{pmatrix},\quad g_{3} = \begin{pmatrix}
        1 &  & & \\&1&&\\&&-1&\\&&&-1
    \end{pmatrix}
\end{equation}
These matrices preserve a Calabi-Yau structure on $\mathbb{C}^{4}$ (as can be seen by diagonalising them) and also the square lattice $\Lambda = (\mathbb{Z}+i\mathbb{Z})^{4}$ and thus we have an action of $G$ on $T^{8} = \mathbb{C}^{4}/\Lambda$ and $T^{8}/G$ is a compact Calabi-Yau orbifold. 

The singular set of this orbifold (with no translations added) is quite large and complicated. It consists of,

\begin{table}[H]
    \centering
    \begin{tabular}{ c | c | c | c }
         & $\# p$ & $\op{Stab}(p)$ & Real Codimension  \\
         \hline
         (1) & 16 & $\langle g_{1}, g_{2},g_{3} \rangle \cong \mathbb{Z}_{2}^{2}\times S_{3}$ & 8\\
         (2) & 64 & $\langle g_{1}g_{2}g_{3}, g_{1}g_{2}^{-1}\rangle \cong D_{6}$ & 8\\
         (3) & 48 & $\langle g_{1}g_{2}^{3}, g_{1}, g_{3}\rangle \cong \mathbb{Z}_{2}^{3}$ & 8\\
         (4) & 16 & $\langle g_{2}^{3},g_{3}\rangle \cong \mathbb{Z}_{2}^{2}$ & 8\\
         (5) & 4 & $\langle g_{1},g_{2}\rangle \cong D_{6}$ & 6\\
         (6) & 4 & $\langle g_{1} g_{3},g_{2}\rangle \cong D_{6}$ & 6\\
         (7) & 16 & $\langle g_{3}g_{1}g_{2}, g_{2}^{2} \rangle\cong S_{3} $ & 6\\
         (8) & 16 & $\langle g_{1}g_{2}^{-1}, g_{1}g_{2} \rangle\cong S_{3} $ & 6\\
         (9) & 12 & $\langle  g_{1}g_{3}, g_{2}^{3}\rangle \cong \mathbb{Z}_{2}^{2}$ & 6\\
         (10) & 16 & $\langle g_{1},g_{3}  \rangle \cong\mathbb{Z}_{2}^{2}$ & 6\\
         (11) & 12 & $\langle g_{1}, g_{1}g_{2}^{3} \rangle \cong \mathbb{Z}_{2}^{2}$ & 6\\
         (12) & 16 & $\langle g_{1}g_{2}^{3}, g_{3} \rangle \cong \mathbb{Z}_{2}^{2}$ & 6\\
         (13) & 1 & $\langle g_{2}\rangle \cong  \mathbb{Z}_{6}$ & 4\\
         (14) & 4 & $\langle g_{2}^{2} \rangle \cong \mathbb{Z}_{3}$ & 4\\
         (15) & 4 & $\langle g_{2}^{3}g_{3}g_{1} \rangle \cong \mathbb{Z}_{2}$ & 4\\
         (16) & 16 & $\langle g_{3} \rangle \cong\mathbb{Z}_{2}$ & 4\\
         (17) & 4 & $\langle g_{2}^{3}\rangle \cong \mathbb{Z}_{2}$ & 4\\
         (18) & 4 & $\langle g_{1} \rangle \cong\mathbb{Z}_{2}$ & 4\\
         (19) & 4 & $\langle g_{1}g_{3}\rangle \cong \mathbb{Z}_{2}$ & 4\\
         (20) & 4 & $\langle g_{1}g_{2}^{3} \rangle \cong\mathbb{Z}_{2}$ & 4
        \end{tabular}
    \caption{Fixed locus of $G$ on $T^{8}/G$}
    \label{tab:my_label}
\end{table}

The singular set of the toroidal orbifold therefore consists of several different components for each row of the table with singularities of the form $\mathbb{C}^{r}/H$ where $r$ is the codimension of the fixed locus and $H$ is the stabiliser group in the 3rd column. All codimension 4 and 6 singularities admit crepant resolutions. There are four types of codimension 8 singularities which arise at the intersections of singularities of lower codimension:
\begin{itemize}
    \item (1) is the intersection of (13), (16) \& (18)
    \item (2) is the intersection of (7) \& (8)
    \item (3) is the intersection of (16), (18) \& (20)
    \item (4) is the intersection of (16) \& (17)
\end{itemize}
It turns out that all of these singularities admit crepant resolutions. The stabliser groups in (1) and (2) appear in \cite{hayashi2017existence} as examples (iv) with $n = 2$ and $m = 6$ or $n = 2$ and $m = 3$ respectively, the stabiliser group in (3) is the Abelian part of example (i) from \cite{hayashi2017existence} with $n=2$ and therefore admits a crepant resolution and (4) is also known to admit a crepant resolution.

One could also classify all the possible space groups with point group $G$. One possible choice is the following,
\begin{equation*}
    (g_{1},t_{1}),\quad (g_{2},\vec{0}),\quad (g_{3},\vec{0})
\end{equation*}
\begin{equation*}
    G' = \langle (g_{1},t_{1}), (g_{2},\vec{0}), (g_{3},\vec{0})\rangle
\end{equation*}
where,
\begin{equation*}
    t_{1} = \frac{1}{2}\begin{pmatrix} 0 \\ 0 \\ 0 \\1+i\end{pmatrix}
\end{equation*}

This greatly simplifies the singular set which is now given by,

\begin{table}[H]
    \centering
    \begin{tabular}{ c | c | c | c }
         & $\# p$ & $\op{Stab}(p)$ & Real Codimension  \\
         \hline
         (1) & 8 & $\langle g_{2},g_{3} \rangle \cong \mathbb{Z}_{2}\times \mathbb{Z}_{6}$ & 8\\
         (2) & 32 & $\langle g_{2}^{2}g_{3}\rangle \cong \mathbb{Z}_{6}$ & 8\\
         (3) & 40 & $\langle g_{2}^{3}, g_{3} \rangle \cong \mathbb{Z}_{2}^{2}$ & 8\\
         (4) & 2 & $\langle g_{2}, g_{1}g_{3} \rangle \cong D_{6}$ & 6\\
         (5) & 8 & $\langle g_{2}g_{1}g_{3}, g_{2}^{2} \rangle \cong S_{3}$ & 6\\
         (6) & 6 & $\langle g_{1}g_{3}, g_{2}^{3} \rangle \cong \mathbb{Z}_{2}^{2}$ & 6\\
         (7) & 1 & $\langle g_{2} \rangle \cong \mathbb{Z}_{6}$ & 4\\
         (8) & 4 & $\langle g_{2}^{2} \rangle \cong \mathbb{Z}_{3}$ & 4\\
         (9) & 2 & $\langle g_{1}g_{3}g_{2}^{3} \rangle \cong \mathbb{Z}_{2}$ & 4\\
         (10) & 8 & $\langle g_{3}\rangle \cong \mathbb{Z}_{2}$ & 4\\
         (11) & 4 & $\langle g_{2}^{3}\rangle \cong \mathbb{Z}_{2}$ & 4\\
         (12) & 2 & $\langle g_{1}g_{3} \rangle \cong \mathbb{Z}_{2}$ & 4
        \end{tabular}
    \caption{Fixed locus of $G'$ on $T^{8}/G'$}
    \label{tab:my_label}
\end{table}
In this case all of the singularities admit crepant resolutions too. The fact that it was relatively easy to find an example of a Calabi-Yau orbifold admitting a crepant resolution, even without adding translations to the group action, is testament to the fact that this problem is much easier in the Calabi-Yau setting than in the hyperK\"ahler one.

\bigskip
\large
\noindent
{\bf {\sf Acknowledgements.}}
\normalsize

We would like to thank G.Bellamy, R. Radhakrishnan, J. Sawon and J. Wi\'sniewski for discussions. We are grateful to Robert Bryant for developing the material of section \ref{sec: Robert's classification} and appendix \ref{sec: trialilty} without which this work would not have been possible. The work of BSA and DB is supported by a grant from the Simons Foundation (\#488569, Bobby Acharya).

\newpage
\bibliographystyle{plain} 

\bibliography{References}

\newpage
\appendix

\section{The Octonions and Triality}

\subsection{The Integral Octonions}\label{sec: int octs}

The octonions are the set of elements,
\begin{equation}
    \mathbb{O} = \left\{x^{0}i_{0} + \sum_{a} x^{a}i_{a} \;\;\middle|\;\; x^{i} \in \mathbb{R}\right\}
\end{equation}
where $i_{0}=1$ and each $i_{a}$ ($a=1,...,7$) is a square root of $-1$. Their algebra can be easily determined by viewing each octonion as a pair of quaternions $a=(a^{1},a^{2})$ and defining the octonion multiplication in terms of the quaternion multiplication in the following way (this is the Cayley-Dickson construction),
\begin{equation}
    a\cdot b = (a^{1},a^{2})\cdot (b^{1},b^{2}) = (a^{1}b^{1}-\bar{b}^{1}a^{2},b^{2}a^{1}+a^{2}\bar{b}^{1})
\end{equation}
we can then define the octonion units $i_{a}$ in terms of the quaternion units $i$, $j$ and $k$ (recalling that $ij=k$ and $i^{2}=j^{2}=k^{2}=-1$),
\begin{equation}
    \left\{i_{0},i_{1},i_{2},i_{3},i_{4},i_{5},i_{6},i_{7}\right\}=\left\{(1,0),(i,0),(j,0),(k,0),(0,1),(0,i),(0,j),(0,k)\right\}
\end{equation}
There are seven quaternionic subalgebras of the form $\{i_{0},i_{a},i_{b},i_{c}\}$ which we will abbreviate as $(0abc)$,
\begin{equation}
    \left\{(0123),(0145),(0246),(0347),(0365),(0761),(0257)\right\}
\end{equation}
Let us adjoin to this the set of complementary indices and denote it $\Sigma^{0}$,
\begin{align}
    \Sigma^{0} = &\left\{(0123),(0145),(0246),(0347),(0365),(0761),(0257)\right.\nonumber\\
    & \left.(4567),(2367),(1357),(1256),(1247),(2345),(1346)\right\}
\end{align}
and define seven other sets $\Sigma^{i}$ by swapping the indices $0$ and $i$ in each element of $\Sigma^{0}$, for example,
\begin{align}
    \Sigma^{1} = &\left\{(0123),(0145),(1246),(1347),(1365),(0761),(1257)\right.\nonumber\\
    & \left.(4567),(2367),(0357),(0256),(0247),(2345),(0346)\right\}
\end{align}
We are now able to define the \emph{integral octonions}. The unit integral octonions consist of the sixteen elements $\pm i_{a}$ ($a=0,..,7)$ and the $16\times 14=224$ elements $\tfrac{1}{2}(\pm i_{a}\pm i_{b}\pm i_{c}\pm i_{d})$ where $(abcd)\in \Sigma^{i}$ (for $i=1,...,7$). The integral octonions are then the lattice generated by integral linear combinations of elements of this set of 240 elements. \textbf{Note:} if one instead tries to define the integral octonions using $(abcd)\in \Sigma^{0}$, the set is not closed under multiplication, this strange fact is known as Kirmse's mistake \cite{conway2013sphere}. We denote the integral octonions by $\mathbb{D}$ and the unit integral octonions by $\bbD^{\times} = \{\ x\in\bbD\ |\ |x|^2 = 1\ \}$. Viewed as a lattice, $\mathbb{D}$ is precisely the $E_{8}$ lattice scaled down by $\sqrt{2}$, so that the roots have squared length $1$ instead of 2. A choice of simple roots could be,
\begin{align}
    &\alpha_{1} = \frac{1}{2}(-i_{2}+i_{3}-i_{4}-i_{5}),\quad \alpha_{2} = \frac{1}{2}(-i_{2}-i_{3}+i_{6}-i_{7}),\quad \alpha_{3} = \frac{1}{2}(i_{2}-i_{3}-i_{6}-i_{7})\nonumber\\
    &\alpha_{4} = \frac{1}{2}(i_{2}+i_{3}+i_{6}+i_{7}),\quad \alpha_{5} = \frac{1}{2}(-i_{2}-i_{3}-i_{6}+i_{7}),\quad \alpha_{6} = \frac{1}{2}(-i_{1}+i_{3}+i_{4}-i_{7})\nonumber\\
    &\alpha_{7} = i_{1},\quad \alpha_{8} = \frac{1}{2}(i_{0}-i_{1}-i_{4}+i_{5});\qquad \alpha_{0} = -1
\end{align}

\subsection{Identities, $\op{Spin}(8)$ and triality} \label{sec: trialilty}

For $p\in\bbO$, we let $L_p:\bbO\to\bbO$ denote left multiplication by~$p$
and let $R_p:\bbO\to\bbO$ denote right multiplication by $p$.  
By Dickson's Theorem, the subalgebra of $\bbO$ 
generated by any two elements is associative. Thus
$$
(L_p)^2 = L_{p^2}\,,\qquad (R_p)^2 = R_{p^2}\,,\qquad\text{and}\qquad
L_p\circ R_p = R_p\circ L_p\,,
$$
although $L_pL_q\not= L_{pq}$, etc.~for general $p,q\in\mathbb{O}$.

Define conjugation $c:\bbO\to\bbO$ to be the linear map 
satisfying $c(1) = 1$ while $c(p)=-p$ when $p$ is purely imaginary.  
Then $p\,\bar p = \bar p\,p = |p|^2$ for all $p\in\bbO$ (where $\bar{p} := c(p)$), 
which implies that, for all $u,v\in\bbO$, 
we have $\la u,v\ra = \tfrac12(u\,\bar v+ v\,\bar u)$.

Another useful fact is the following:  
When $\rho_u:\bbO\to\bbO$ is reflection in the hyperplane 
perpendicular to $u\not=0$, we have
$$
\rho_u(v) =  v  - 2\frac{\la u,v\ra}{|u|^2}\,u
= v - \frac{(u\,\bar v+ v\,\bar u)u}{|u|^2} = -\frac{u\,\bar v\,u}{|u|^2},
$$
since $(v\,\bar u) u = |u|^2 v$.  When $|u|^2 = 1$, 
since $c = -\rho_1$, we have the identities
$$
\rho_u\circ \rho_1 = L_{u}\circ R_{u}\qquad\text{and}\qquad
\rho_1 \circ\rho_u = L_{\bar u}\circ R_{\bar u}\,,
$$
and $c\circ\rho_u\circ c = \rho_{\bar u}$, which will be useful below.

For any pair $x,y\in\bbO$, we have the identities $\la x,yz\ra 
= \la x\,\bar z,y\ra  = \la \bar y\,x,z\ra$, which imply 
$$
\la x,y\ra 
= \Re(x\,\bar y) 
= \Re(y\,\bar x),
$$
which implies that $\Re(xy) = \Re(yx)$.  

Moreover, even though the octonions are not associative, we have
$$
\Re\bigl((xy)z\bigr) = \Re\bigl(x(yz)\bigr) 
= \Re\bigl((zx)y\bigr),
$$
so that the expression $\Re(xyz)$ is unambiguous.  (However, note
that, generally, $\Re(xyz)$ is not equal to $\Re(xzy)$.)

Two special facts are noteworthy:  First there is L. E. Dickson's theorem that any subalgebra of $\bbO$ generated by two elements is associative.  Second, there is Moufang's identity: $x(yz)x = (xy)(zx)$.

The group $\Spin(8)\subset\SO(8)^3$ 
is defined to be the set of triples $(g_1,g_2,g_3)$ 
such that, for all $x,y,z\in\bbO$,
$$
\Re\bigl(g_1(x)g_2(y)g_3(z)\bigr) = \Re(xyz).
$$
This group is connected and simply-connected and has the property 
that each of the factor projections~$\pi_i:\Spin(8)\to\SO(8)$ 
is a non-trivial double cover.  
Moreover, the nontrivial element of the $\bbZ_2$-kernel of~$\pi_i$ 
is $(\epsilon_1\,I_8,\epsilon_2\,I_8,\epsilon_3\,I_8)$, 
where $\epsilon_i = 1$ and $\epsilon_j=-1$ for $j\not=i$.   
These $3$ elements, together with~$(I_8,I_8,I_8)$, 
constitute the center of~$\Spin(8)$.

Since $\Re(xyz) = \Re(yzx) 
= \Re(\bar z\,\bar y\,\bar x)$, 
the maps $\alpha(g_1,g_2,g_3) = (cg_3c,cg_2c,cg_1c)$ 
and $\beta(g_1,g_2,g_3) = (g_2,g_3,g_1)$ 
are automorphisms of~$\Spin(8)$. 
They generate a group of order~$6$ that is isomorphic to~$S_3$.  
This group acts effectively on the center of~$\Spin(8)$
and represents the group of \emph{outer} automorphisms of~$\Spin(8)$.  

Setting $\PSO(8) = \SO(8)/\{\pm I_8\}$, 
and letting $q:\SO(8)\to\PSO(8)$ be the double cover, 
there is are automorphisms $\bar\alpha$ and $\bar\beta$
of~$\PSO(8)$ 
such that $q{\circ}\pi_i\circ\beta = \bar\beta{\circ}q{\circ}\pi_i$
and $q{\circ}\pi_i\circ\alpha = \bar\alpha{\circ}q{\circ}\pi_i$.

Note that, by the Moufang identity, if $|u|^2 = 1$, then
$$
\Re\bigl((\bar u\,x\,\bar u)(uy)(zu)\bigr) 
= \la u\bar x u,(uy)(zu)\ra 
= \la u\bar x u, u(yz)u\ra = \la \bar x , yz\ra
= \Re(xyz),
$$
so that $(L_u{\circ}R_u, L_{\bar u}, R_{\bar u})$ is an element of~$\Spin(8)$.

In fact, $\Spin(8)$ is generated by the $7$-sphere
$$
\Sigma 
= \{\,(L_v{\circ}R_v, L_{\bar v}, R_{\bar v})\,|\,|v|=1\,\}\subset\Spin(8).
$$
This follows from the classical fact that $\Or(8)$ is generated by reflections in hyperplanes through the origin in $\bbO=\bbR^8$, i.e., by $\{\,\rho_u\,|\,|u|=1\,\}$, while $\SO(8)$ is generated by products of two such reflections, in particular, by $\{\,\rho_u{\circ}\rho_1\,|\,|u|=1\,\}$ while $\rho_u{\circ}\rho_1 = L_u{\circ}R_u$.  

Let $W\subset\bbO$ be a subspace, and define $\SO(W)$ to be the subgroup 
of~$\SO(\bbO)$ that is the identity on $W^\perp$.  Then $\SO(W)$ 
is generated by an even number of the reflections~$\rho_u$ such that $u\in W$.
Define $\Spin(W) = \pi_1^{-1}\bigl(\SO(W)\bigr)\subset\Spin(8)$.  
When $\dim W\ge 2$, the group $\Spin(W)$ 
is a nontrivial (connected) double cover of $\SO(W)$. 
Moreover, when $\dim W < 8$, we have $-I_8\not\in\SO(W)$, 
implying that the images $\pi_2\bigl(\Spin(W)\bigr)$ 
and $\pi_3\bigl(\Spin(W)\bigr)$ are still non-trivial double covers
of $\SO(W)$.  

In particular, if $\dim W = 7$, then $\pi_i\bigl(\Spin(W))\simeq\Spin(7)$
for $i=2,3$; if $\dim W = 6$, then $\pi_i\bigl(\Spin(W))\simeq\Spin(6)=\SU(4)$
for $i=2,3$; and if $\dim W = 5$, then $\pi_i\bigl(\Spin(W))\simeq\Spin(5)=\Sp(2)$ for $i=2,3$.

Our main interest in this is that, when $\Gamma$ 
is a subgroup of $\SO(W)$, the corresponding groups 
$\Gamma_i = \pi_i\bigl(\pi_1^{-1}(\Gamma)\bigr)$ for $i=2,3$ 
will be subgroups of $\pi_i\bigl(\Spin(W)\bigr)\simeq\Spin(W)$.

\section{Classifying Translations}\label{sec:class trans}

In this appendix, we classify the possible translations that can be added to the group actions corresponding to the blue highlighted and starred rows in the tables of section \ref{sec: list of subgroups}. In the language of crystallography, each of these rows corresponds to a point group and we classify all the possible space groups, up to affine equivalence. To do this, we used GAP \cite{GAP4} to implement an algorithm due to Zassenhaus \cite{zassenhaus1948algorithmus} and described by Brown in \cite{brown1969algorithm}. We then study the singular sets of the arising orbifolds to determine whether or not they admit symplectic resolutions. Firstly, we review the basics on space groups (we found the notes \cite{souvignier2008group} useful) and outline the algorithm (following \cite{brown1969algorithm}). Then we summarise our results.

\subsection{Determining all space groups given a point group}

\vspace{1em}
\noindent{\emph{Preliminaries on space groups:}}
\vspace{1em}

Take $N\in \op{GL}(n,\mathbb{R})$ and $t\in \mathbb{R}^{n}$. We denote an affine transformation of $\mathbb{R}^{n}$ by $(N, t)$, acting as $(N,t)\cdot x = Nx+t$. Products and inverses are thus given by,
\begin{equation}
    (N,t)(N',t') = (NN',t+Nt'),\qquad (N,t)^{-1} = (N^{-1},-N^{-1}t)
\end{equation}
Define the \emph{Euclidean group} $\mathcal{E}_{n}$ as the group of all affine mappings $(N,t)$ such that $N\in \op{O}(n)$. There is an obvious group homomorphism,
\begin{align*}
    \pi:\;\; \mathcal{E}_{n}&\to \op{O}(n)\\
     (N,t) &\mapsto N
\end{align*}
A subgroup $G\subset\mathcal{E}_{n}$ is called a \emph{space group} if it contains a maximal lattice in $\mathbb{R}^{n}$ as a subgroup (maximal meaning the basis of the lattice spans $\mathbb{R}^{n})$. In other words, the kernel $T := \{(\mathbb{1},t)\in G\}$ of the mapping $\pi|_{G}:G\to \op{O}(n)$ defines a maximal lattice,
\begin{equation}
    L_{T} := \left\{ \sum_{i=1}^{n} c_{i}t_{i}\mid c_{i}\in \mathbb{Z},\; (\mathbb{1},t)\in G \right\} 
\end{equation}
which we will call the \emph{translation lattice}. The image $\pi(G)$ is called the \emph{point group} of $G$ and is isomorphic to $G/T$. Note that, importantly, $G$ usually does not contain a subgroup isomorphic to its point group. This occurs when, for example, $G$ contains an element $(N,t)$ with $N\neq \mathbb{1}$ such that $t$ is not in the translation lattice. 

Since $T$ is the kernel of $\pi$, it is a normal subgroup of $G$ and therefore the point group is a subgroup of the automorphism group $\pi(G)\subset \op{Aut}(L_{T})$. As a consequence, $\pi(G)$ is always finite since $\op{Aut}(L_{T})$ is. A matrix $N\in \op{O}(n)$ is an automorphism of $L_{T}$ if $N$ and $N^{-1}$ send a basis of $L_{T}$ to another basis of $L_{T}$, this can be summarised as,
\begin{equation}
    \op{Aut}(L_{T}) = \left\{N\in \op{O}(n)\mid N M_{gen} = M_{gen}C, \;\; C\in \op{GL}(n,\mathbb{Z})\right\}
\end{equation}
where $M_{gen}$ is the matrix whose columns form a basis of $L_{T}$. Note that the matrices $C$ form an integral representation of $\op{Aut}(L)$.

Since the point group acts on and preserves the lattice $L_{T}$, it is natural to write the space group and its point group with respect to the lattice basis $\{t_{i}\}$ rather than the standard basis $\{e_{i}\}$ of $\mathbb{R}^{n}$. In fact, this is necessary to facilitate implementing the algorithm we will describe later on a computer. Transforming to the lattice basis amounts to conjugating the whole space group by $M_{gen}$,
\begin{equation}
    (N,t)\mapsto (N',t') = \left(M_{gen}^{-1}\;N \; M_{gen},\; M_{gen}^{-1}\;t \right),\quad \forall (N,t)\in G
\end{equation}
This has the following consequences:
\begin{itemize}
    \item The translation lattice now becomes $\mathbb{Z}^{n}$
    \item The point group matrices $N'$ are no longer orthogonal and now lie in $\op{GL}(n,\mathbb{Z})$. Instead, they preserve the intersection form $A = M_{gen}^{T}M_{gen}$, i.e. $(N')^{T}A N' = A$. This is how we retain the knowledge of our original lattice.
\end{itemize}

\vspace{2em}
\noindent{\emph{Defining a space group given a point group:}}
\vspace{1em}

Let $P$ be a subgroup of $\op{GL}(n,\mathbb{Z})$ preserving the intersection form $A$ of some lattice in $\mathbb{R}^{n}$ and let $\{N_{1},...,N_{p}\}$ be a choice of generators of $P$. A set of vectors $\tau = \{t_{a},t_{b},...\}$ in $\mathbb{R}^{n}$, one vector for each element $a\in P$, is called a \emph{vector system} for $P$ if and only if,
\begin{equation}\label{eq: vector systems rule}
    t_{a b} = t_{a} + a\;t_{b} \mod \mathbb{Z}^{n},\quad \forall a,b\in P
\end{equation}
If we are given an abstract finite presentation of $P$
\begin{equation}
    \langle N_{1},...,N_{p}\mid R_{1},...,R_{r}\rangle
\end{equation}
where each $R_{i}$ is a word in the $N_{i}$ that equals $\mathbb{1}$, then a vector system is completely determined modulo $\mathbb{Z}^{n}$ by the set of vectors $\{t_{N_{1}},t_{N_{2}},...,t_{N_{p}}\}$ such that $t_{R_{i}} = \vec{0}\mod \mathbb{Z}^{n}$ (where $t_{R_{i}}$ is evaluated using the rule \eqref{eq: vector systems rule}).

A space group (in the lattice basis) is then defined by a point group $P\subset \op{GL}(n,\mathbb{Z})$ preserving the intersection form $A$ of some lattice and a vector system $\tau$. The generators of the space group are given by $(N_{i},t_{N_{i}})$. We consider two space groups $G$ and $G'$ to be the same if they are \emph{affinely equivalent}, meaning
\begin{equation}
    G' = (X,\eta)^{-1}\;G\; (X,\eta)
\end{equation}
for some $(X,\eta)\in \mathcal{E}_{n}$. It is true that there are only finitely many space groups, up to affine equivalence, in each dimension and that two space groups are abstractly isomorphic if and only if they are affinely equivalent \cite{bieberbach1911bewegungsgruppen, bieberbach1912gruppen, bieberbach1912minkowskische}.

The problem is then: given a point group, determine a representative of each affine equivalence class of space groups with that point group.

To this end, it is useful to define two vector systems $\tau' = \{t'_{a}\}$ and $\tau = \{t_{a}\}$ for a point group $P$ to be \emph{equivalent} if there exists an 
\begin{equation}
    X\in N_{\op{GL}(n,\mathbb{Z})}(P) = \{X\in\op{GL}(n,\mathbb{Z})\mid X^{-1}PX = P \}
\end{equation}
and $\eta \in \mathbb{R}^{n}$ such that
\begin{equation}\label{eq: equivalent vec sys}
    t_{a}' = X\; t_{X^{-1}\,a\,X} + (\mathbb{1}-a)\eta
\end{equation}
(this formula arises from conjugating $(a,t_{a})$ by $(X,\eta)$). Let $\mathcal{V}_{P}$ be a set of representatives of the classes of equivalent vector systems for $P$. Zassenhaus \cite{zassenhaus1948algorithmus} showed that each $n$-dimensional space group with point group $P$ is affinely equivalent to one and only one of the space groups built from $P$ and some $\tau\in \mathcal{V}_{P}$. This means that each affine equivalence class has a unique distinguished representative given by a space group built from $P$ and some $\tau\in \mathcal{V}_{P}$.

Thus, to determine a representative of all affine equivalence classes of space groups with a given point group $P$, we need only determine $\mathcal{V}_{P}$.

\vspace{2em}
\noindent{\emph{Brief outline of the algorithm:}}
\vspace{1em}

We first define two vector systems to be \emph{strongly equivalent} if they satisfy $\eqref{eq: equivalent vec sys}$ with $X = \mathbb{1}$. This amounts to a shift of the origin. We denote the set of representatives of strongly equivalent vector systems by $\tilde{\mathcal{V}}_{P}$

The algorithm to determine $\mathcal{V}_{P}$ (the set of representatives of classes of equivalent vector systems) for a given point group $P$ goes as follows (see \cite{brown1969algorithm} for details):
\begin{enumerate}
    \item Determine a finite presentation of $P$ given by generators $N_{1},...,N_{p}$ and relations $R_{1},...,R_{r}$ and determine all the vector systems of $P$ (i.e. all $\tau = \{t_{N_{i}}\}$ such that $t_{R_{i}}=\vec{0}\mod \mathbb{Z}^{n}$).
    \item Find representatives of the classes of strongly equivalent vector systems.
    \item Compute the normaliser of $P$ in $\op{GL}(n,\mathbb{Z})$
    \begin{equation*}
        N_{\op{GL}(n,\mathbb{Z})}(P) = \{X\in\op{GL}(n,\mathbb{Z})\mid X^{-1}PX = P \}
    \end{equation*}
    and reduce the list of representatives of classes of strongly equivalent vector systems to a list of representatives of classes of equivalent vector systems.
\end{enumerate}

\noindent For solving steps 1 and 2, observe that we can write $t_{R_{i}}$ in terms of the $t_{N_{i}}$ as,
\begin{equation}
    t_{R_{i}} = \sum_{a=1}^{p}B(i,j)t_{N_{j}},\quad j=1,...,r
\end{equation}
where each coefficient $B(i,j)$ is an $n\times n$ matrix. These coefficient matrices can be efficiently computed using GAP. One can combine these $pr$ distinct matrices into a single $pn\times rn$ matrix $B$ and form the $nk\times 1$ vector $\hat{t}$ by concatenating all the $t_{N_{i}}$ so that $t_{R_{i}}=\vec{0}$ for $i=1,...,r$ is equivalent to the matrix equation $B\;\hat{t}=\vec{0} \mod \mathbb{Z}^{n}$. Then diagonalise $B$ over the integers to obtain $D = PBQ$ for $P,Q\in \op{GL}(n,\mathbb{Z})$. As explained in \cite{brown1969algorithm}, one can then simply read off the representatives of the strong equivalence classes: if $D=\op{diag}(d_{1},d_{2},...,d_{q},0,0,...,0)$ then the representatives of the strong equivalence classes are the $d_{1}\cdot d_{2}\cdot ... \cdot d_{q}$ vectors,
\begin{equation}
    Q\begin{pmatrix}
        t_{1}/d_{1}\\t_{2}/d_{2}\\ \vdots\\t_{q}/d_{q}\\0\\ \vdots\\0
    \end{pmatrix},\quad t_{i}\in \mathbb{Z}
\end{equation}

If one is only interested in computing the possible singular sets that arise from quotienting $T^{8}$ by these space groups then one can stop here. Indeed we had to do this for some examples as computing the normaliser of some point groups proved to take too long to justify (we will indicate which examples in the next subsection). However for all other examples we proceeded with step 3 for completeness. One can compute the normaliser in GAP using the function \texttt{NormalizerInGLnZ}. One can then compute the action of each generator of the normaliser using similar methods to the computation of $B$. Each generator acts as some permutation on the set $\tilde{\mathcal{V}}_{P}$. One must find the orbits under the group generated by these permutations and choose a representative of each orbit and the algorithm is complete.

For studying the singular sets of the orbifolds, we found the GAP crystallography package CRYST \cite{Cryst} to be an invaluable tool, in particular the function \texttt{WyckoffPositions} for computing the fixpoints of a group action modulo a lattice.

\subsection{Our results}

There are 15 blue rows and 8 starred rows in the tables of section \ref{sec: list of subgroups}. We take the group in each row and change it to the lattice basis (as per the previous subsection) and then follow the algorithm to compute all possible space groups with that point group. We summarise below how many different space groups there are for each case and whether or not the corresponding orbifold has NRS-singularities. In the Github repository \url{https://github.com/D-A-Baldwin/HyperKahler-8mfds} we list, for each point group $H$, all possible choices of space group $\hat{H}$ and the singular set of $\mathbb{R}^{8}/\hat{H}$. It is also quite possible that some of these groups are conjugate in $\op{GL}(n,\mathbb{Z})$ but we did not check this.

The only resolvable case occurs in row 24 of table 3, for the case with trivial translations. This orbifold is identical to the one considered in \S 6 of \cite{donten201781} where they prove it has the betti numbers $b_{2} = 23$, $b_{4} = 276$ which are identical to those of $\op{Hilb}^{2}(K3)$. Later in \cite{donten2017very}, it was proven that this resolution is in fact diffeomorphic to $\op{Hilb}^{2}(K3)$.

\subsubsection{Table \ref{tab:table 1}:}

\begin{table}[H]
    \centering
    {
    \renewcommand{\arraystretch}{1.5}
    \begin{tabular}{c|c|c|c|c}
        Row & Group & $\# \tilde{\mathcal{V}}_{P}$ & $\# \mathcal{V}_{P}$ & NRS-singularities?  \\
        \hline\hline
        1 & $S_{3}$ & 1 & 1 & Yes (e.g. $\mathbb{C}_{4}/\mathbb{Z}_{3}$) \\
        6 & $D_{4}\cong \mathbb{Z}_{2}\wr\mathbb{Z}_{2}$ & 1 & 1 & Yes (e.g. $\mathbb{C}_{4}/\mathbb{Z}_{4}$) \\
        7 & $D_{4}\cong \mathbb{Z}_{2}\wr\mathbb{Z}_{2}$ & 1 & 1 & Identical singular set to row 6 \\
        8 & $D_{4}\cong \mathbb{Z}_{2}\wr\mathbb{Z}_{2}$ & 1 & 1 & Identical singular set to row 6 \\
        12 & $D_{4}\cong \mathbb{Z}_{2}\wr\mathbb{Z}_{2}$ & 4 & 2 & Yes for all cases (e.g. $\mathbb{C}^{4}/\mathbb{Z}_{4}$) \\
        44 & $Q_{8}\circ D_{4}$ & 2 & 2 & Yes for all cases (e.g. $\mathbb{C}^{4}/\mathbb{Z}_{4}$) \\
        49 & $\mathbb{Z}_{4}\wr \mathbb{Z}_{2}$ &  2 & 2 & Yes for all cases (e.g. $\mathbb{C}^{4}/\mathbb{Z}_{4}$) \\
        50 & $\mathbb{Z}_{4}\wr \mathbb{Z}_{2}$ & 2 & 2 & Identical singular set to row 49 \\
        52 & $\mathbb{Z}_{4}\wr \mathbb{Z}_{2}$ & 2 & 2 & Identical singular set to row 49 \\
        74 & $Q_{8}\wr \mathbb{Z}_{2}$ &  8 & \xmark & Yes for all cases (e.g. $\mathbb{C}^{4}/\mathbb{Z}_{4}$)\\
    \end{tabular}
    }
    \caption{The results of classifying the translations of the blue rows of Table \ref{tab:table 1}. A `\xmark'  in the 4-th column means we didn't compute the integral normaliser and therefore didn't compute $\mathcal{V}_{P}$.}
    \label{tab:blue rows of table 1}
\end{table}

\begin{table}[H]
    \centering
    {
    \renewcommand{\arraystretch}{1.5}
    \begin{tabular}{c|c|c|c|c}
        Row & Group & $\# \tilde{\mathcal{V}}_{P}$ & $\# \mathcal{V}_{P}$ & NRS-singularities?  \\
        \hline\hline
         71 & $Q_{8}\cdot D_{6}$ & 4 & \xmark & Yes for all cases (e.g. $\mathbb{C}^{4}/\mathbb{Z}_{4}$) \\
         72 & $Q_{8}\cdot D_{6}$ & 4 & \xmark & Identical singular set to row 71 \\
         73 & $Q_{8}\cdot D_{6}$ & 4 & \xmark & Identical singular set to row 71 \\
        
    \end{tabular}
    }
    \caption{The results of classifying the translations of the starred rows of Table \ref{tab:table 1}. A `\xmark'  in the 4-th column means we didn't compute the integral normaliser and therefore didn't compute $\mathcal{V}_{P}$.}
    \label{tab:starred rows of table 1}
\end{table}

\subsubsection{Table \ref{tab:table 2}:}

\begin{table}[H]
    \centering
    {
    \renewcommand{\arraystretch}{1.5}
    \begin{tabular}{c|c|c|c|c}
        Row & Group & $\# \tilde{\mathcal{V}}_{P}$ & $\# \mathcal{V}_{P}$ & NRS-singularities?  \\
        \hline\hline
         8 & $\mathbb{T}$ & 1 & 1 & Yes (e.g. $\mathbb{C}^{4}/\mathbb{Z}_{4}$) \\
        
    \end{tabular}
    }
    \caption{The results of classifying the translations of the blue rows of Table \ref{tab:table 2}. A `\xmark'  in the 4-th column means we didn't compute the integral normaliser and therefore didn't compute $\mathcal{V}_{P}$.}
    \label{tab:blue rows of table 2}
\end{table}

\begin{table}[H]
    \centering
    {
    \renewcommand{\arraystretch}{1.5}
    \begin{tabular}{c|c|c|c|c}
        Row & Group & $\# \tilde{\mathcal{V}}_{P}$ & $\# \mathcal{V}_{P}$ & NRS-singularities?  \\
        \hline\hline
         14 & $\mathbb{I}$ & 1 & \xmark & Yes (e.g. $\mathbb{C}^{4}/\mathbb{Z}_{4}$) \\
         15 & $\op{SmallGroup}(720,409)$ & 4 & \xmark & Yes (e.g. $\mathbb{C}^{4}/\mathbb{Z}_{3}$) \\
    \end{tabular}
    }
    \caption{The results of classifying the translations of the starred rows of Table \ref{tab:table 2}. A `\xmark'  in the 4-th column means we didn't compute the integral normaliser and therefore didn't compute $\mathcal{V}_{P}$.}
    \label{tab:starred rows of table 2}
\end{table}

\subsubsection{Table \ref{tab:table 3}:}

\begin{table}[H]
    \centering
    {
    \renewcommand{\arraystretch}{1.5}
    \begin{tabular}{c|c|c|c|c}
        Row & Group & $\# \tilde{\mathcal{V}}_{P}$ & $\# \mathcal{V}_{P}$ & NRS-singularities?  \\
        \hline\hline
        2 & $\mathbb{D}_{4}\cong \mathbb{Z}_{2}\wr \mathbb{Z}_{2}$ & 16 & 3 & Yes for all cases (e.g. $\mathbb{C}^{4}/\mathbb{Z}_{2}$ or $\mathbb{C}^{4}/\mathbb{Z}_{2}$) \\
        24 & $Q_{8}\circ D_{4}$ & 4096 & 26 & No for trivial translation\tablefootnote{This group is conjugate in $\op{GL}(8,\mathbb{Z})$ to the group considered in \S 6 of \cite{donten201781}, thus for the case of trivial translations we get an identical orbifold to theirs which has a resolution diffeomorphic to $\op{Hilb}^{2}(K3)$ \cite{donten2017very}.}, \\
        &&&& yes for all other cases (e.g. $\mathbb{C}^{4}/\mathbb{Z}_{2}$ or $\mathbb{C}^{4}/\mathbb{Z}_{2}$) \\
        27 & $\mathbb{Z}_{4}\wr \mathbb{Z}_{2}$ &  16 & 5 & Yes for all cases (e.g. $\mathbb{C}^{4}/\mathbb{Z}_{4}$) \\
        42 & $Q_{8}\wr \mathbb{Z}_{2}$ & 64 & \xmark & Yes for all cases (e.g. $\mathbb{C}^{4}/\mathbb{Z}_{4}$) \\
        
    \end{tabular}
    }
    \caption{The results of classifying the translations of the blue rows of Table \ref{tab:table 3}. A `\xmark'  in the 4-th column means we didn't compute the integral normaliser and therefore didn't compute $\mathcal{V}_{P}$.}
    \label{tab:blue rows of table 3}
\end{table}

\begin{table}[H]
    \centering
    {
    \renewcommand{\arraystretch}{1.5}
    \begin{tabular}{c|c|c|c|c}
        Row & Group & $\# \tilde{\mathcal{V}}_{P}$ & $\# \mathcal{V}_{P}$ & NRS-singularities?  \\
        \hline\hline
         45 & $D_{4}\cdot S_{4}$ & 4 & \xmark & Yes (e.g. $\mathbb{C}^{4}/\mathbb{Z}_{4}$) \\
         46 & $(Q_{8}^{2}\rtimes \mathbb{Z}_{2})\rtimes \mathbb{Z}_{3}$ & 4 & \xmark & Yes (e.g. $\mathbb{C}^{4}/\mathbb{Z}_{4}$) \\
         48 & $\op{SmallGroup}(1920,241003)$ & 16 & \xmark & Yes (e.g. $\mathbb{C}^{4}/\mathbb{Z}_{4}$) \\
    \end{tabular}
    }
    \caption{The results of classifying the translations of the starred rows of Table \ref{tab:table 3}. A `\xmark'  in the 4-th column means we didn't compute the integral normaliser and therefore didn't compute $\mathcal{V}_{P}$.}
    \label{tab:starred rows of table 3}
\end{table}

\section{The HyperK\"ahler Structure Preserved by the Groups $G_{i}$}\label{sec: basis change mats}
Here we provide for each group $G_{i}$ a basis change matrix $B^{(i)}$, such that the matrices,
\begin{equation}
    T'^{(i)}_{j} = (B^{(i)})^{-1}\;T^{(i)}_{j}B^{(i)},\quad j=1,2,3,4
\end{equation}
preserve the standard hyperK\"ahler structure on $\mathbb{C}^{4}$ (where we choose complex coordinates $z^{i} = x^{2i-1} + ix^{2i}$ in terms of the coordinates $x^{i}$ on $\mathbb{R}^{8}$),
\begin{equation}
    \omega_{1} = \frac{i}{2}\sum_{i=1}^{4}dz^{i}\wedge d\bar{z}^{i},\quad \omega_{2}+i\omega_{3} = dz^{1}\wedge dz^{2}+dz^{3}\wedge dz^{4}
\end{equation}
Note that after this basis change, the lattice preserved by the matrices has generator matrix,
\begin{equation}
    M_{\text{gen}}' = (B^{(i)})^{-1}\,M_{\text{gen}}
\end{equation}
The basis change matrices are,
\begin{align*}
    B^{(1)} &= \begin{pmatrix}
        1 & 0 & 0 & 0 & 0 & 0 & 0 & 0 \\
 0 & 1 & 0 & 0 & 0 & 0 & 0 & 0 \\
 0 & 0 & 0 & 0 & -1 & 0 & 0 & 0 \\
 0 & 0 & 0 & 0 & 0 & 1 & 0 & 0 \\
 0 & 0 & 1 & 0 & 0 & 0 & 0 & 0 \\
 0 & 0 & 0 & 1 & 0 & 0 & 0 & 0 \\
 0 & 0 & 0 & 0 & 0 & 0 & 1 & 0 \\
 0 & 0 & 0 & 0 & 0 & 0 & 0 & 1
    \end{pmatrix}\\
    \\
    B^{(2)} &= \begin{pmatrix}
        -1/\sqrt{6} & 0 & 0 & -1/\sqrt{2} & 0 & 0 & 0 &
   1/\sqrt{3} \\
 0 & 0 & 0 & 0 & -1 & 0 & 0 & 0 \\
 0 & 0 & -\sqrt{2/3} & 0 & 0 & -1/\sqrt{3} & 0 & 0 \\
 0 & -1/\sqrt{2} & 1/\sqrt{6} & 0 & 0 & -1/\sqrt{3} & 0
   & 0 \\
 0 & 0 & 0 & 0 & 0 & 0 & 1 & 0 \\
 0 & 1/\sqrt{2} & 1/\sqrt{6} & 0 & 0 & -1/\sqrt{3} & 0 &
   0 \\
 1/\sqrt{6} & 0 & 0 & -1/\sqrt{2} & 0 & 0 & 0 &
   -1/\sqrt{3} \\
 \sqrt{2/3} & 0 & 0 & 0 & 0 & 0 & 0 & 1/\sqrt{3}
    \end{pmatrix}\\
    \\
    B^{(3)} &= \begin{pmatrix}
        1 & 0 & 0 & 0 & 0 & 0 & 0 & 0 \\
 0 & 1 & 0 & 0 & 0 & 0 & 0 & 0 \\
 0 & 0 & 0 & 0 & 0 & -1 & 0 & 0 \\
 0 & 0 & 0 & 0 & 1 & 0 & 0 & 0 \\
 0 & 0 & 1 & 0 & 0 & 0 & 0 & 0 \\
 0 & 0 & 0 & 1 & 0 & 0 & 0 & 0 \\
 0 & 0 & 0 & 0 & 0 & 0 & -1 & 0 \\
 0 & 0 & 0 & 0 & 0 & 0 & 0 & 1
    \end{pmatrix}
\end{align*}

\end{document}